\documentclass[fleqn,usenatbib]{mnras}

\usepackage{newtxtext,newtxmath}

\usepackage[T1]{fontenc}

\DeclareRobustCommand{\VAN}[3]{#2}
\let\VANthebibliography\thebibliography
\def\thebibliography{\DeclareRobustCommand{\VAN}[3]{##3}\VANthebibliography}

\usepackage{graphicx}	
\usepackage{amsmath}	
\usepackage{newtxtext,newtxmath}	
\usepackage{multirow}
\usepackage{tabularx}
\usepackage{arydshln}

\title[Dust and gas production of comet 67P]{A quantitative description of comet 67P's dust and gas production remains enigmatic}

\author[D. Bischoff et al.]{
D. Bischoff\thanks{E-mail: d.bischoff@tu-bs.de}$^{1}$,
C. Schuckart$^{1}$,
N. Attree$^{1}$,
B. Gundlach$^{2}$,
J. Blum$^{1}$\\
$^{1}$Institut f\"ur Geophysik und Extraterrestrische Physik, Technische Universit\"at Braunschweig, Mendelssohnstr. 3, 38106 Braunschweig, Germany\\
$^{2}$Institut f\"ur Planetologie, Universität Münster, Wilhelm-Klemm-Str. 10, 48149 Münster, Germany\\
}

\date{Accepted XXX. Received YYY; in original form ZZZ}

\pubyear{2023}

\begin{document}
\label{firstpage}
\pagerange{\pageref{firstpage}--\pageref{lastpage}}
\maketitle

\begin{abstract}
The mechanism of dust emission from a cometary nucleus is still an open question and thermophysical models have problems reproducing outgassing and dust productions rates simultaneously. In this study, we investigate the capabilities of a rather simple thermophysical model to match observations from Rosetta instruments at comet 67P/Churyumov-Gerasimenko and the influence of model variations. We assume a macro-porous surface structure composed of pebbles and investigate the influence of different model assumptions. Besides the scenario in which dust layers are ejected when the vapour pressure overcomes the tensile strength, we use artificial ejection mechanisms, depending on ice-depletion of layers. We find that dust activity following the pressure criterion is only possible for reduced tensile strength values or reduced gas diffusivity and is inconsistent with observed outgassing rates, because activity is driven by CO$_2$. Only when we assume that dust activity is triggered when the layer is completely depleted in H$_2$O, the ratio of CO$_2$ to H$_2$O outgassing rates is in the expected order of magnitude. However, the dust-to-H$_2$O ratio is never reproduced. Only with decreased gas diffusivity, the slope of the H$_2$O outgassing rate is matched, however absolute values are too low. To investigate maximum reachable pressures, we adapted our model equivalent to a gas-impermeable dust structure. Here, pressures exceeding the tensile strength by orders of magnitude are possible. Maximum activity distances of $3.1 \,\mathrm{au}$, $8.2 \,\mathrm{au}$, and $74 \,\mathrm{au}$ were estimated for H$_2$O-, CO$_2$-, and CO-driven activity of $1 \,\mathrm{cm}$-sized dust, respectively. In conclusion, the mechanism behind dust emission remains unclear. 
\end{abstract}
\begin{keywords}
methods: numerical -- comets: general -- radiation mechanisms: thermal -- conduction
\end{keywords}



\section{Introduction}
\label{sec:introduction}
Comets are characterised by their dust and gas activity observable in their comae and tails. However, the physical processes resulting in activity are not well understood. It is clear, that the sublimation of volatiles is key for cometary activity, but how it can eject dust particles or chunks thereof is part of current theoretical, numerical and experimental research. The so-called cohesion bottleneck describes the unsolved problem that the low pressure of the escaping gas flow must overcome the cohesion of the material plus the gravitational force to eject dusty material \citep{Keller.1993,Jewitt.2019}. Gravity is usually negligible on comets. The cohesion, which can be expressed by the tensile strength, depends on material properties \citep{Bischoff.2020} and the structure in terms of e.g., granularity and volume filling factor \citep{Skorov.2012}. Micrometer-sized dust has a higher tensile strength than larger dust grains due to the larger contact surface area. Compressed material also possesses a higher tensile strength than more porous samples. Agglomerates of micrometer-sized dust have a reduced tensile strength due to the reduced contact area of the agglomerates. Regarding comets, such agglomerates or “pebbles” may play an important role and are widely expected to be the building blocks of comets formed in protoplanetary disks \citep{Blum.2017,Ciarniello.2021,Ciarniello.2022,Fulle.2022b,Blum.2022}. 

The pressure that can be reached by a sublimating ice is determined by the temperature, the structure of the overlying material and the ice-material itself. Carbon monoxide is extremely volatile and sublimates efficiently at temperatures around $30$ to $40 \,\mathrm{K}$, whereas carbon dioxide needs temperatures around $100$ to $120 \,\mathrm{K}$. Water ice starts to sublimate effectively around $180 \,\mathrm{K}$. Therefore, these ices will trigger activity at different distances from the Sun or different depths under the cometary surface. Observationally, activity of comets was found at distances even beyond $20 \,\mathrm{au}$ \citep[][]{Jewitt.2019,Jewitt.2021b,Farnham.2021}. It is clear that this activity cannot be driven by the outgassing of water ice. Besides the sublimation, which consumes energy, also phase transition that release energy might play an important role in dust activity, such as the transition from amorphous to crystalline water ice \citep{Prialnik.2008}. Other volatiles might be trapped in amorphous water ice and will be released when the phase change occurs \citep{BarNun.1985,BarNun.1987}. This transition is significant for temperatures $>100\,\mathrm{K}$. For micro-granular water ice, sintering becomes significant when temperatures for efficient sublimation are reached \citep{Gundlach.2018}. Sintering describes the growing necks between ice grains in contact, changing the heat conductivity, which might play a role for cometary activity. For a pressure build-up, the structure of the overlying material is crucial. In less permeable systems, higher pressure can be reached compared to systems in which the gas can easily escape into space. 

\begin{figure}
    \centering
    \includegraphics[width=\columnwidth]{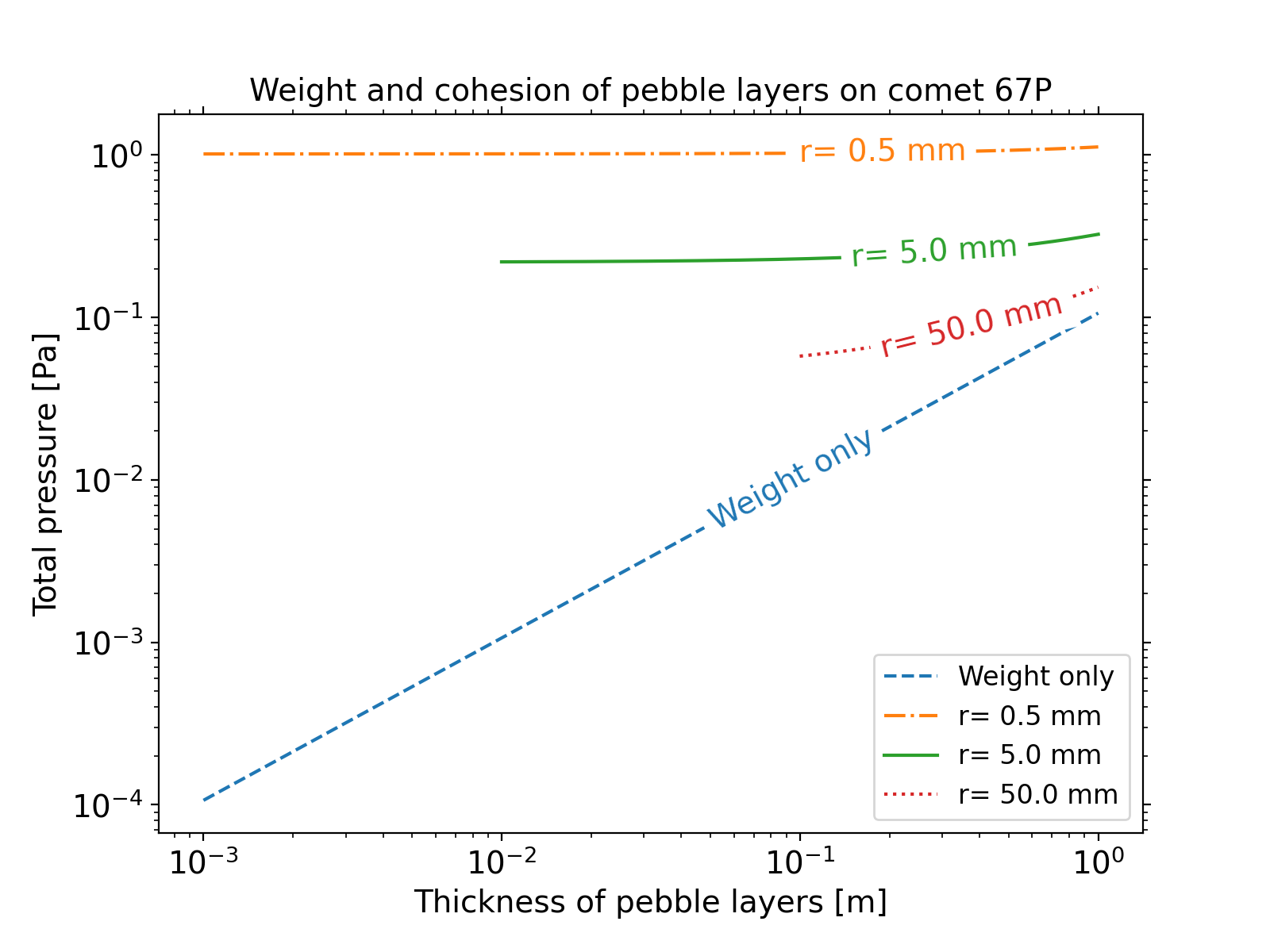}
    \caption{Sum of tensile strength of pebble layers (Equation \ref{eq:pebble_tensile_strength}) and layer weight as a function of layer thickness for pebbles with radii $R= 0.5 \,\mathrm{mm}$ (orange, dashed-dotted), $R= 5 \,\mathrm{mm}$ (green, solid) and $R= 50 \,\mathrm{mm}$ (red, dotted), assuming a mass density of $532 \,\mathrm{kg/m^3}$ and an acceleration of gravity of $2 \times 10^{-4} \,\mathrm{m/s}$. The lines start at a layer thickness corresponding to the pebble diameter. The blue dashed line represents the pressure resulting from weight of the layer only.}
    \label{fig:Mass_Cohesion_Pressure}
\end{figure}

Measurements of the tensile strength of cometary surfaces resulted in a range of several orders of magnitude and depend on the observed length scale. An overview of measurements of tensile and compressive strengths for cometary materials can be found in \citet{Biele.2022b}. The analysis of a cliff collapse on comet 67P/Churyumov-Gerasimenko (hereafter comet 67P) derived tensile strength values on the order of $1 \,\mathrm{Pa}$ \citep{Attree.2018}. From the observation of breakups of cometary meteors with a size of millimetres in Earth’s atmosphere, the tensile strength on small length scale can be estimated. It was found that they possess on the order of $1 - 10 \,\mathrm{kPa}$ \citep{Blum.2014}. For piles of pebbles with radius $R$, \citet{Skorov.2012} derived theoretically tensile strengths of
\begin{equation}
    \sigma(R) =  \sigma_{0} \, \Phi_{\mathrm{pack}}  \left(\frac{R}{1 \,\mathrm{mm}} \right)^{-2/3} ,
    \label{eq:pebble_tensile_strength}
\end{equation}
with $\sigma_{0}=1.6\,\mathrm{Pa}$ and $\Phi_{\mathrm{pack}}$ being the volume filling factor of the pebble packing. Thus, for pebble sizes of millimetres to centimetres \citep{Zsom.2010,Lorek.2018}, low tensile strengths $\lesssim 1 \, \mathrm{Pa}$ are expected and were experimentally confirmed \citep{Blum.2014,Brisset.2016}. Figure \ref{fig:Mass_Cohesion_Pressure} shows the expected stress required to release pebble layers as a function of depth below the surface of comet 67P for different pebble radii $R$. It includes the tensile strength from Equation \ref{eq:pebble_tensile_strength} and the weight of each layer assuming a density of $532 \,\mathrm{kg/m^3}$ and an acceleration of gravity of $2 \times 10^{-4} \,\mathrm{m/s}$. The stress resulting from the weight only is plotted for comparison. For pebbles with radii up to centimetre size, the weight only plays a role for depths larger than several decimetres. 

The tensile strength inside a pebble is comparable to the values found for meteors, but depends on the material composition. In the laboratory, it was measured for silica \citep[$\sim \mathrm{kPa}$,][]{Meisner.2012,Gundlach.2018b}, water ice \citep[$\sim \mathrm{kPa}$,][]{Gundlach.2018b}, carbon dioxide \citep[tens to hunders of kPa,][]{Fritscher.2022} and several organic materials \citep[hundreds of Pascals up to tens of kPa,][]{Bischoff.2020}.

Due to these tensile strength values, it might be necessary to reduce the cohesion for dust emission. Thermal processing, shear deformation \citep{Matonti.2019} and the formation of cracks might be of importance for cometary activity \citep{Jewitt.2019}. In general, a wide spectrum of geological phenomena have been observed on comets \citep{Sunshine.2016}. On comet 67P, several cracks on the metre scale were observed \citep{ElMaarry.2015}. They could be attributed to sintering processes, combined with thermal fracturing \citep{Auger.2018,Hofner.2017}. \citet{Attree.2018b} modelled the stresses induced by temperature changes and found that stresses can penetrate to typical depths of $\sim 25 \,\mathrm{cm}$. These and other geological effects might reduce the cohesion, which could solve the cohesion bottleneck problem. However, further investigations of such processes are needed. 

In the past decades, many researchers worked on thermophysical models, which try to simulate the conditions on comets \citep[see e.g.,][]{Keller.2017,Prialnik.2017,Fulle.2019,Fulle.2020,Fulle.2020b,Hu.2019,Gundlach.2020,Davidsson.2021c, Davidsson.2021, Davidsson.2022}. Several approaches were used and different aspects were considered in the various models. However, in all cases simplifications or restrictions had to be applied, due to the complexity and unknown aspects of cometary surfaces. 

The modelling approach of \citet{Davidsson.2021} was able to fit the outgassing rates of water and carbon dioxide measured at comet 67P with the ROSINA instrument. They used a dust-to-ice mass ratio of 1 to 2 and a CO$_2$ abundance of 11 to 32 per cent of the total ice mass. Their fitting parameters were the pore length and pore radius, and they assumed that the outgassing of water emits the same mass of dust without other criteria for the ejection of dust. In their model, the depth of the sublimation front of the two ices evolved and resulted in a dust mantle thickness of typically $\lesssim 2\,\mathrm{cm}$. CO$_2$ sublimated near aphelion at depths between $\sim 1.9\,\mathrm{m}$ (southern hemisphere) and $\sim 3.8 \,\mathrm{m}$ (northern hemisphere). To fit the data, \citet{Davidsson.2021} needed a shift in the fit parameters around perihelion, which was explained with airfall material that evolves during the aphelion passage, and material compaction due to CO$_2$ sublimation.  

\citet{Macher.2019} used a 3D thermal model of two regions on 67P to investigate the influence of tangential heat flow and compared their results to MIRO data. They found deviations between a 1D and a 3D model of up to $30 \,\mathrm{K}$ for the first $5 \,\mathrm{cm}$ below the surface. In general, in case of 67P the complex shape has an important influence. \citet{Macher.2019} were able to match the MIRO sub-mm-wavelength results only with a discrepancy to the mm-wavelength observations.

Another model was proposed by \citet{Fulle.2019,Fulle.2020,Fulle.2020b}, which assumes a pebble structure in which the pebbles themselves possess a substructure of smaller aggregates. In such configurations, higher pressures can be reached resulting in ongoing dust activity when erosion dominates over dehydration. They found that erosion and water-vapour flux do not depend on the dust-to-ice mass ratio of the nucleus. The dust-to-ice mass ratio effects the dehydration rate only. The water production rates fit nicely the measurements from comet 67P \citep{Ciarniello.2021} for relatively high dust-to-ice mass ratios $>5$ \citep{Fulle.2020b}. 

\citet{Skorov.2020} found that the observed steep increase of gas production can be explained by an increase of the active-area fraction near the south pole. Also, the polar day at comet 67P enables sublimation in deeper layers, due to the deeper penetration of the heat wave. In general, it should be noted that the complexity of comet 67P, due to its irregular shape and large obliquity, results in many effects, which often need to be ignored for the modelling. 
\citet{Skorov.2021,Skorov.2022} found that a variation in porosity of the dust layer results in a change of gas production depending on the layer thickness. Macroscopic cavities and cracks do not influence the gas production significantly.

\citet{Gundlach.2020} compared model results to observations from comet 67P's south-pole region at perihelion and found that CO$_2$ sublimation drives the ejection of large dust chunks, whereas water sublimation ejects small dust aggregates. 
The work of \citet{Bischoff.2021} did not include ices in their thermophysical model, because they concentrated on surface temperature. However, they showed that one can distinguish between a micro-porous and a macro-porous surface structure by observing the surface temperature at sunrise for varying insolation at day. 

In this paper, we first present in Section \ref{sec:TPM} the thermophysical model and the modifications compared to the earlier work described above. In Section \ref{sec:results_outgassing}, we compare the different approaches for modelling the outgassing rate of the sublimating ices and different implementations of dust activity. To investigate the maximum pressure reachable in comets, we present different approaches in Section \ref{sec:results:pressure}, including an adaptation of our model equivalent to a gas-impermeable dust structure. We discuss our results in comparison to findings from the Rosetta mission and other models in Section \ref{sec:discussion} and conclude with an outlook in Section \ref{sec:conclusions}.

\section{Thermophysical model}
\label{sec:TPM}
\begin{figure*}
    \centering
    \includegraphics[width=\textwidth]{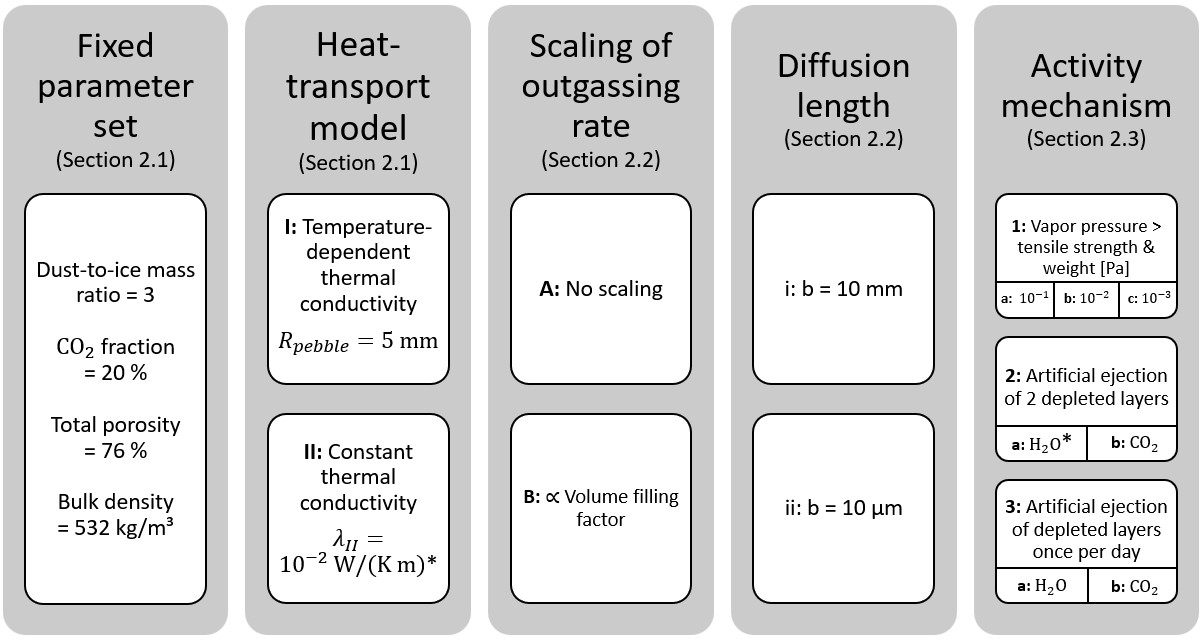}
    \caption{Overview of model variations. Due to run time capacity constraints, we did not run all possible combinations of variations, but selected for some aspects the most promising combinations. (*): We also performed simulation without a day-night cycle and therefore permanent illumination. These were performed for the case I.A.i.2a and additionally for two macro-porous cases with adapted constant thermal conductivities ($10^{-4}\,\mathrm{W/(K \,m)}$ and $10^{-1}\,\mathrm{W/(K \,m)}$). }
    \label{fig:Overview}
\end{figure*}

In the following, we describe the thermophysical model and the variations we are investigating in this work. An overview of the fixed parameter set and the different cases are presented in Figure \ref{fig:Overview}. The different combinations of model variations are denoted with abbreviations, which are explained in this section and can be found in Figure  \ref{fig:Overview}. 

\subsection{Basics}
\label{sec:TPM:Basis}
The one-dimensional thermophysical model developed by \citet{Gundlach.2012, Gundlach.2020} and \citet{Bischoff.2021} is the basis for our model. We model one square metre of a cometary surface and subsurface down to a depth of one to two metres. To the model of \citet{Bischoff.2021}, which assumed only refractory material, water and carbon dioxide ice was added. The mass fractions of dust, H$_2$O and CO$_2$ add up to unity, 
\begin{equation}
    f_{m,\mathrm{dust}} + f_{m,\mathrm{H_{2}O}} + f_{m,\mathrm{CO_2}} = 1.
    \label{eq:mass_fraction1}
\end{equation}
We solve the heat-transfer equation for time $t$ and depth $z$
\begin{equation}
    \rho(z) c(z) \frac{dT(z,t)}{dt} = \frac{d}{dz} \left[ \lambda(T(z),z) \frac{dT(z,t)}{dz} \right] + Q(T(z),z)
    \label{eq:heat_transport_eq}
\end{equation}
using the finite difference method and the forward difference scheme with constant $dz$ and $dt$. The thermal conductivity $\lambda(T(z),z)$ depends on the given model scenario, which is explained in the following paragraphs. It is either given by Equation \ref{eq:pebble_conductivity} or by Equation \ref{eq:const_conductivity}. The local density $\rho(z)$ and heat capacity $c(z)$ of a numerical layer depend on the local composition, i.e. the local volume fraction of dust, H$_2$O, CO$_2$ and vacuum, combined with the respective densities and heat capacities of the materials. The heat capacities of the three materials depend on the local temperature and are given in Equations \ref{eq:heat_capacity_water} to \ref{eq:heat_capacity_dust}. Therefore, these values are updated in each time step. The source and sink term $Q$ accounts for the sublimation and re-sublimation of ice molecules and is the sum of $Q_X$, given in Equation \ref{eq:latent_heat}, with $X$ representing the species H$_2$O or CO$_2$. For the surface layer, the insolation and the outgoing thermal radiation by the Stefan-Boltzmann law are taken into account. We assume an initial temperature of $50\,\mathrm{K}$ throughout the body, which is also the lower boundary condition. For simplicity and run time constrains, we do not model the gas flow explicitly, but use the approach presented in \citet{Gundlach.2020} of using a damping factor for the outgassing resulting from dust layers above the sublimation front (for details, see Section \ref{sec:TPM:ScalingOutgassing}). 

In \citet{Bischoff.2021}, the macro-porous case, which assumes a surface structure of pebbles, is compared to the micro-porous case, comparable to a homogeneous micrometre-sized dust structure. This comparison is adapted in this study. Here, we compare the influence of temperature-dependent thermal conductivity from radiative heat transport, which is denoted with I in the following scenarios (see Figure \ref{fig:Overview}), to a case of a constant thermal conductivity, denoted with II. It should be noted here that with this adaptation, we do not further investigate the influence of the surface structure, but only the influence of the temperature dependency. The basic assumption behind the temperature-dependent model is a pebble structure of the material and therefore that the total thermal conductivity is the sum of the phononic heat transport through the contacting solid grains and the photonic heat transport by radiation through the voids between the pebbles. Details of that part of the model can be found in \citet{Bischoff.2021}. For simplicity, we assume that the network thermal conductivity is temperature-independent, which is compared to the strong temperature-dependency of the radiative conductivity a reasonable simplification. Thus, the total heat conductivity reads
\begin{equation}
    \lambda_{\mathrm{I}} = \lambda_{\mathrm{net}}(R) + \lambda_{\mathrm{rad}}(R,T) = \lambda_{\mathrm{net}}(R) + p_{\mathrm{rad}}(R)\, \left( \frac{T}{\mathrm{1~K}} \right)^3. 
    \label{eq:pebble_conductivity}
\end{equation}
The derivation of the parameters $\lambda_{\mathrm{net}}(R)$ and $p_{\mathrm{rad}}(R)$ can be found in Appendix \ref{App:TPM}. $\lambda_{\mathrm{net}}(R)$ is not constant, due to its dependency on the mass fractions. However, this influence is generally small and negligible for high temperatures when radiation dominates. For this study here, we chose one fixed pebble radius of $R = 5 \,\mathrm{mm}$, which is in agreement with the findings of earlier work \citep{Blum.2017,Burger.2022}. 

For comparison, in the second scenario II, we use a constant, i.e. temperature-independent, thermal conductivity and for simplicity, we chose a single value of
\begin{equation}
    \lambda_{\mathrm{II}} = \lambda_{\mathrm{net}}(r) = 10^{-2} \, \mathrm{W \, m^{-1} K^{-1}},
    \label{eq:const_conductivity}
\end{equation}
which is comparable to values reported for comet 67P \citep{Marshall.2018,Spohn.2015}.
We used these two cases of heat-transport laws to investigate the different model variations, which are presented in Sections \ref{sec:results_sub:macro} and \ref{sec:results_sub:micro}. For both cases, we also assume a fixed set of other input parameters, with a dust-to-ice mass ratio of 3:1 (including all ices), a CO$_2$ fraction of 20 \% of the total ice mass and a total porosity of 0.76. These values are educated guesses based on the observations made at comet 67P. The bulk density of $532 \,\mathrm{kg/m^3}$, which is used here, is known with a high accuracy \citep{Jorda.2016}. For the assumed pebble structure, which has two stages of porosity, it is distributed via an intra-pebble porosity of 0.4 and an inter-pebble porosity of the packing of 0.6 \citep{Blum.2017}. 

Compared to the model of \citet{Gundlach.2020}, several additions were made. We added the day-night cycle and the possibility to simulate an elliptical orbit. Here, we model the cometary surface at the equator of comet 67P for the duration of 1000 cometary days, beginning at a heliocentric distance of $4.11 \,\mathrm{au}$ and ending at perihelion. This part of the orbit is most interesting, as the activity evolves. We chose to model only the inbound branch for simplicity and the influence of the comets obliquity, resulting in seasonal effect. For example, \citet{Davidsson.2021} needed a shift in the fitting parameters of their thermophysical model around perihelion to match the inbound and outbound branches. Although the inbound branch is also influenced by seasonal effects, we decided to only use this part of the orbit. For the specific heat capacity and the latent heat, temperature-dependent values were used. These relations can be found, beside all other used physical properties, in Appendix \ref{App:TPM}. The volume-filling factor evolves with time when the ices sublimate and layers get depleted. Beside these additions, some changes in the assumptions were made. The thermal conductivity in the temperature-dependent models is now calculated as an interface parameter, which means that it is not calculated with the mean temperature of the corresponding layer, but with the mean temperature of the two layers that exchange energy. This has an effect for high temperature gradients, which occurs at the upper centimetres.

\subsection{Scaling of outgassing rate}
\label{sec:TPM:ScalingOutgassing}
Another important addition to the model is the implementation of two different approaches for the calculation of the outgassing rate $j_{\mathrm{leave}}$ of each layer. The Hertz-Knudsen formula is adapted by adding a factor for the damping by depleted layers above the sublimation front $\eta$ and a factor for the fraction of sublimating surface area $\alpha$,
\begin{equation}
    j_{\mathrm{leave},X} = p_{\mathrm{sat},X}(T) \, \sqrt{\frac{m_X}{2 \pi k T}} \, \eta \, \alpha .
    \label{eq:outgassing}
\end{equation}
$X$ denotes the ice species, so either H$_2$O or CO$_2$. Here, $p_{\mathrm{sat},X}(T)$, $m_X$, $k$ and $T$ are the saturation pressure, the mass of the molecule $X$, the Stefan-Boltzmann constant and the temperature, respectively. \citet{Gundlach.2020} assumed a sublimating dust-ice interface of one square meter, regardless of the ice abundances. Scenarios with this assumption, where $\alpha = 1$, are denoted with A in our study. Besides this, we also use a second approach. The surface area of the porous medium is larger than the modelled cross section of one square meter. However, experiments have shown that a sample of granular water ice sublimates at the same rate as a solid sample (Christopher Kreuzig, personal communication). Therefore, we exclude scaling factors $\alpha > 1$. Due to the mixture of ices with dust, the volume of the sublimating ice compared to a pure-ice situation is reduced. Hence, our second approach for $\alpha$, which is denoted with B in the scenario nomenclature, is to equate it with the volume filling factor $\Phi$, i.e. $\alpha = \Phi$. Due to the balance of sublimation and re-sublimation, it is likely that the dust (and inactive ice) is covered by a thin layer of ice. It became apparent that this part of the thermophysical model influences the resulting pressures. However, the physical correctness of the different implementations is unclear, which is the reason why we used both approaches and compare their results.

The reduction of the outgassing rate by overlying dust layers at the depth $z$, $\eta$, uses the parameter $b$, which equals the number of aggregate layers required to reduce the outgassing rate by a factor of 2,
\begin{equation}
    \eta = \left( 1 + \frac{z}{b} \right)^{-1} .
    \label{eq:eta}
\end{equation}
This description was made by \citet{Gundlach.2011} and is also used in \citet{Gundlach.2012} and \citet{Gundlach.2020}. The parameter $b$ was experimentally determined in \citet{Gundlach.2011} to be $b = 7$ aggregate radii, however other work resulted in lower values \citep{Gundlach.2020,Schweighart.2021}. To reduce the number of free parameters, we initially chose to set $b = 10 \, \mathrm{mm} \equiv 1$ aggregate diameter, which is denoted with i in the scenario nomenclature (see Figure \ref{fig:Overview}). We take the same value in the temperature-independent case. However, this parameter can be crucial because it directly influences outgassing rates and pressures. For comet 67P, fluffy fractal particles have been observed in the coma and \citet{Fulle.2017} argued, that they might have been stored during formation in the voids of the pebble structure and were released due to activity. In the case of such fluffy dust in the voids, the gas diffusivity would be drastically decreased, but the radiative heat transport through the voids would not be influenced, due to the smallness of the dust particles. To investigate the influence of such a scenario, we ran selected model cases with a parameter $b = 10 \, \mathrm{\mu m}$, which are denoted with ii (see Figure \ref{fig:Overview}). To investigate the influence of the surface structure, as done in \citet{Bischoff.2021} who compared a macro-porous pebble structure to a micro-porous structure, a further adaptation of the diffusion length is necessary. The diffusion through micro-porous micrometre-sized dust grains is less efficient than through a pebble structure with large voids, even if they are filled with fractals. However, this investigation is beyond the scope of this work and might be done in the future. 

The pressure in each layer at the depth $z$, where ice $X$ is abundant, reads \citep{Gundlach.2020}:
\begin{equation}
    p_X = p_{\mathrm{sat},X}(T) \left( 1 - \left( 1 + \frac{z}{b} \right)^{-1} \right)
    \label{eq:pressure_sat}
\end{equation}
In layers without any ice we assume that the pressure vanishes. 

With the outgassing flow of molecules and the latent heat of species $X$, the resulting energy increase or decrease is calculated through
\begin{equation}
    Q_X = \Lambda_X \, (j_{\mathrm{inward},X} - j_{\mathrm{leave},X}) \, dt \, \mathrm{[J/m^2]}.
    \label{eq:latent_heat}
\end{equation}
It is assumed that half of the outgassed mass escapes into space, whereas the other half diffuse into the interior, resulting in the $j_{\mathrm{inward},X}$ term. Due to the lower temperature at lower depths, the molecules will condense on the dust and ice surface. However, for night times, when the temperature gradient inverts, this assumption is wrong. However, due to the lower outgassing rates and the direct sublimation of the condensed material in the morning, the influence should be minor. With a simple model of molecules travelling through a pebble-structured matrix, which possesses a temperature gradient and a resulting condensation coefficient, the distribution of condensing molecules was calculated (Thilo Glißmann, personal communication). As the exact distribution has not a large influence on the simulation results, we do not go into detail about this modelling here. However, we assume that 29\% of the inwards diffusing molecules condense in the first lower layer, 35\% in the second layer, 24\% in the third layer and 12\% in the fourth layer, respectively.

\subsection{Activity mechanism}
\label{sec:TPM:Activity}

\citet{Gundlach.2020} simulated dust activity by the ejection of layers when the vapour pressure of one of the ices exceeded the tensile strength of the layer, which was assumed constant or depth-dependent. We also used this approach, which we denote with 1 in the scenario nomenclature (see Figure \ref{fig:Overview}), with a variation of the constant threshold values. The value for case 1a, $10^{-1} \,\mathrm{Pa}$, corresponds roughly to the tensile strength of pebbles with a radius of $R = 5\,\mathrm{mm}$, which is assumed in all cases (see Figure \ref{fig:Mass_Cohesion_Pressure} and Equation \ref{eq:pebble_tensile_strength}). When a pebble lies on a bed of other pebbles, there will be at least 3 monomer contacts between the pebbles. In this case, these monomer contacts must be broken to detach the pebble. For such monomer contacts, a binding force of $10^{-7} \,\mathrm{N}$ applies \citep{Heim.1999}, and this results in a tensile strength on the order of $4 \times 10^{-3} \,\mathrm{Pa}$. For simplicity, we chose for case 1b a value of $10^{-2} \,\mathrm{Pa}$. The lowest value in case 1c, $10^{-3} \,\mathrm{Pa}$, corresponds to no cohesion between the particles so that only the weight of the pebble is relevant here (see Fig \ref{fig:Mass_Cohesion_Pressure}). Therefore, this is the lowest boundary, which must be overcome in any case. The thickness of the ejected dust layers corresponds to the depth in which the threshold was reached and is therefore not fixed by the model.

However, because it is unclear whether the tensile strength might be reduced by some effects on the cometary surface and subsurface and other effects trigger the ejection of dust, we also used two further approaches. These artificially assume ejections independent of the vapour pressure. The mechanism of scenario 2 (see Figure \ref{fig:Overview}) ejects a fixed number of two layers as soon as they are depleted of the respective ices (case 2a for H$_2$O ice, case 2b for CO$_2$ ice). Therefore, the ejected dust has always a thickness of $1 \,\mathrm{cm}$.

The other mechanism, which is denoted with 3 (see Figure \ref{fig:Overview}), ejects with a fixed frequency all layers of dust that are devoid of the respective ice (case 3a for H$_2$O ice, case 3b for CO$_2$ ice). The frequency was set to once per cometary day and due to this approach, the thickness of the ejected dust is not fixed. We also ran models with this activity mechanism added by an energy criterion, in which an insolated energy corresponding to the perihelion situation must be accumulated before ejections are possible, but there was no different to the case without such a criterion. Therefore, we did not included it in our model variations. 

In all cases, at least two layers must be ejected, which corresponds to one pebble diameter ($1 \,\mathrm{cm}$). 

\section{How to reach measured outgassing rates?}
\label{sec:results_outgassing}
In this Section, our results from the model variations are presented and compared to measured data from the Rosetta mission to comet 67P. The outgassing rates of water and carbon dioxide from \citet[][]{Lauter.2020} are used and re-scaled to one square metre of cometary surface and per second (units for the outgassing rates are, thus, $\mathrm{kg \, s^{-1} \, m^{-2}} $). Here, we use the upper limit for the surface area of 67P of $51.7 \times 10^6 \,\mathrm{m^2}$ \citep{Preusker.2017,Jorda.2016} so that the re-scaling results in a lower limit. To enhance the clarity with reducing the scattering of the measured data, we fitted functions to them and plotted a resulting range (see Figure \ref{fig:Rosina_Fit}; blue, hatched: H$_2$O, green: CO$_2$). The derivation of theses ranges is described in Appendix \ref{App:Rosetta_Data}. Data regarding the dust production rate for different heliocentric distances are rare. \citet{Fulle.2016c} investigated the change of the size distribution along the orbit. They found a total dust loss rate of $60 \pm 10 \,\mathrm{kg/s}$ at $2.2 \,\mathrm{au}$, $70 \pm 30 \,\mathrm{kg/s}$ at $2.1 \,\mathrm{au}$ and $(1.7 \pm 0.9) \times 10^4 \,\mathrm{kg/s}$ around perihelion including all mass bins. \citet{Marschall.2020b} found a dust production rate of $500 \,\mathrm{kg/s}$ around perihelion. In the following plots, we used these ranges scaled to one square metre, analogous to the outgassing rate, marked in grey-hatched.   

\begin{figure*}
    \centering
    \begin{minipage}[b]{.48\linewidth}
    \includegraphics[width=\columnwidth]{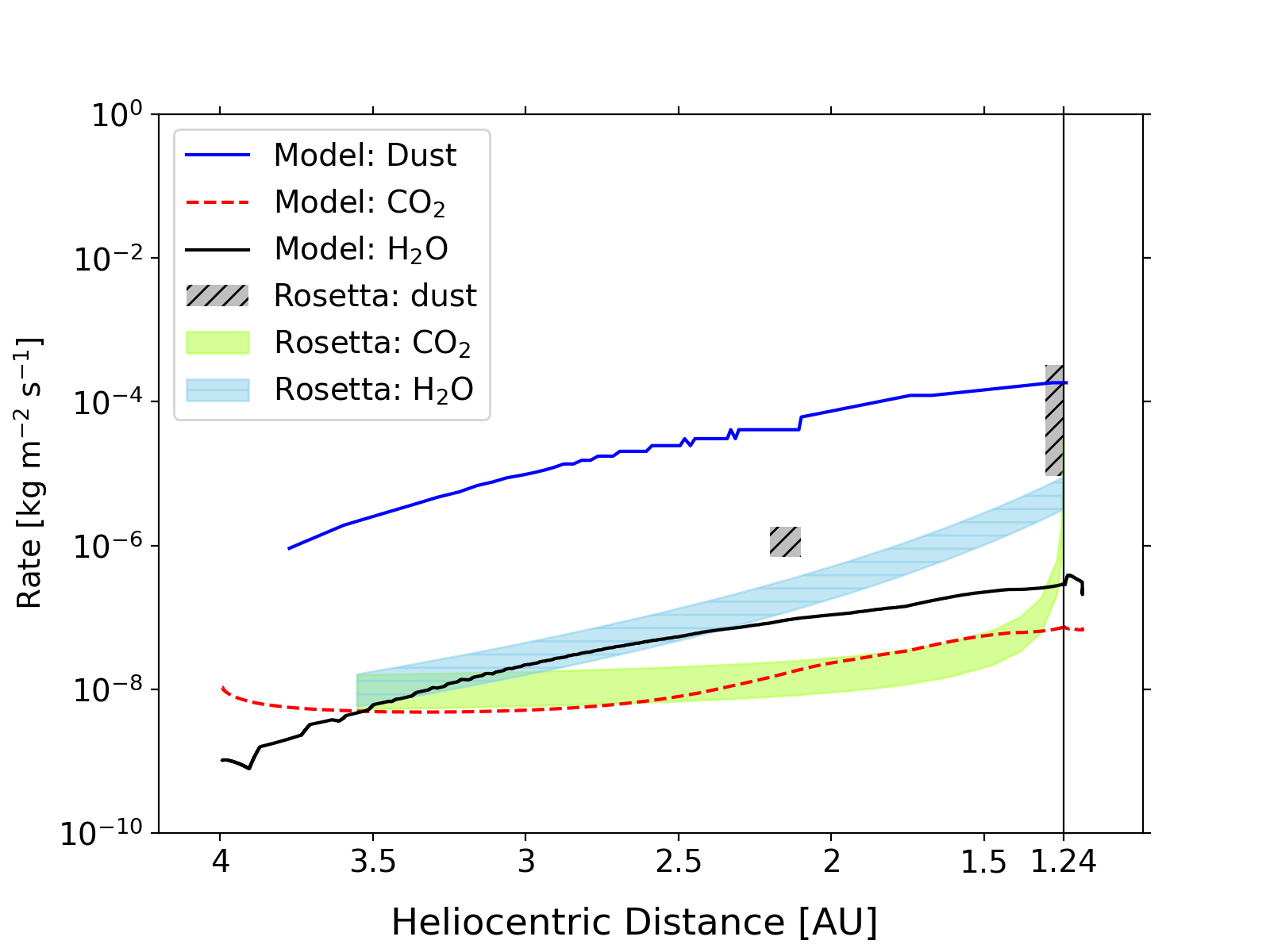}
    \end{minipage}
    \hspace{.03\linewidth}
    \begin{minipage}[b]{.48\linewidth}
    \centering
    \includegraphics[width=\columnwidth]{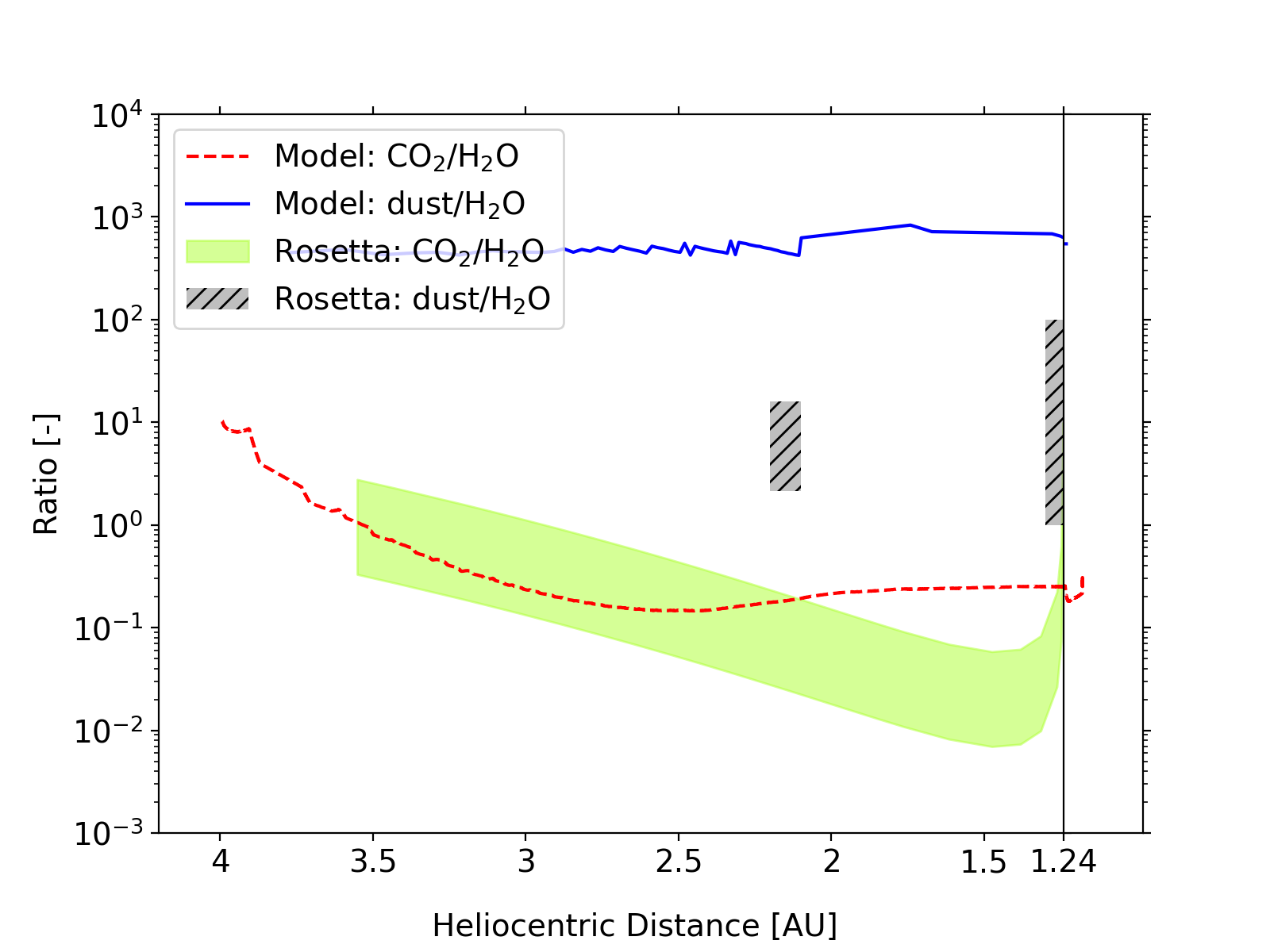}
    \end{minipage}
    \caption{Results for case I.A.i.3a (temperature-dependent thermal conductivity, no scaling, $b=10\,\mathrm{mm}$, H$_2$O-depleted layers once per day). Left: Outgassing- and dust-productions rates for the inbound orbit compared to Rosetta measurements at 67P. Right: Ratio between CO$_2$- and H$_2$O-outgassing rates and ratio between dust-production and H$_2$O-outgassing rates for the inbound orbit, compared to the Rosetta measurements.} 
    \label{fig:results_I_3a_A} 
\end{figure*}

\begin{figure*}
    \centering
    \begin{minipage}[b]{.48\linewidth}
    \includegraphics[width=\columnwidth]{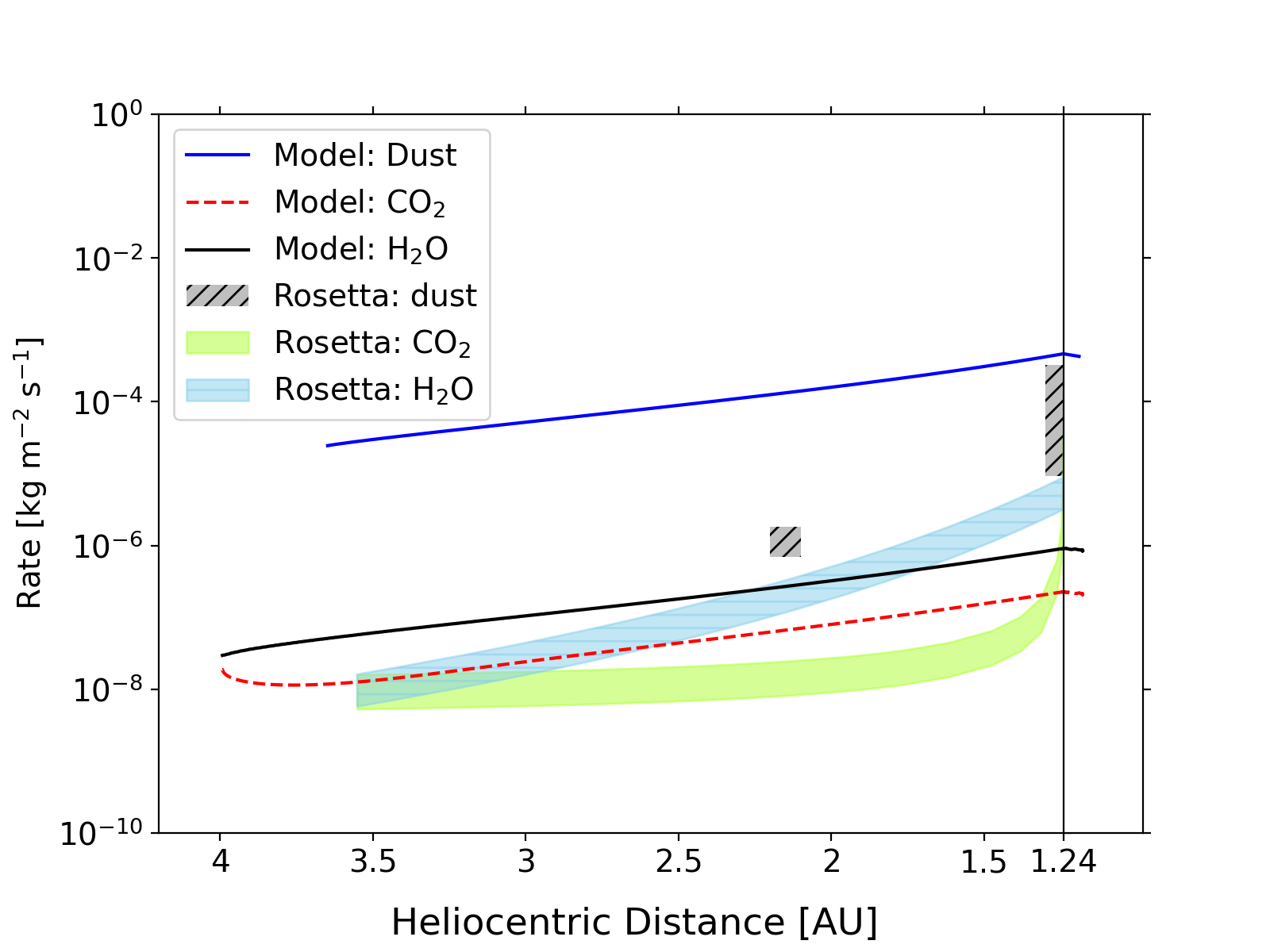}
    \end{minipage}
    \hspace{.03\linewidth}
    \begin{minipage}[b]{.48\linewidth}
    \centering
    \includegraphics[width=\columnwidth]{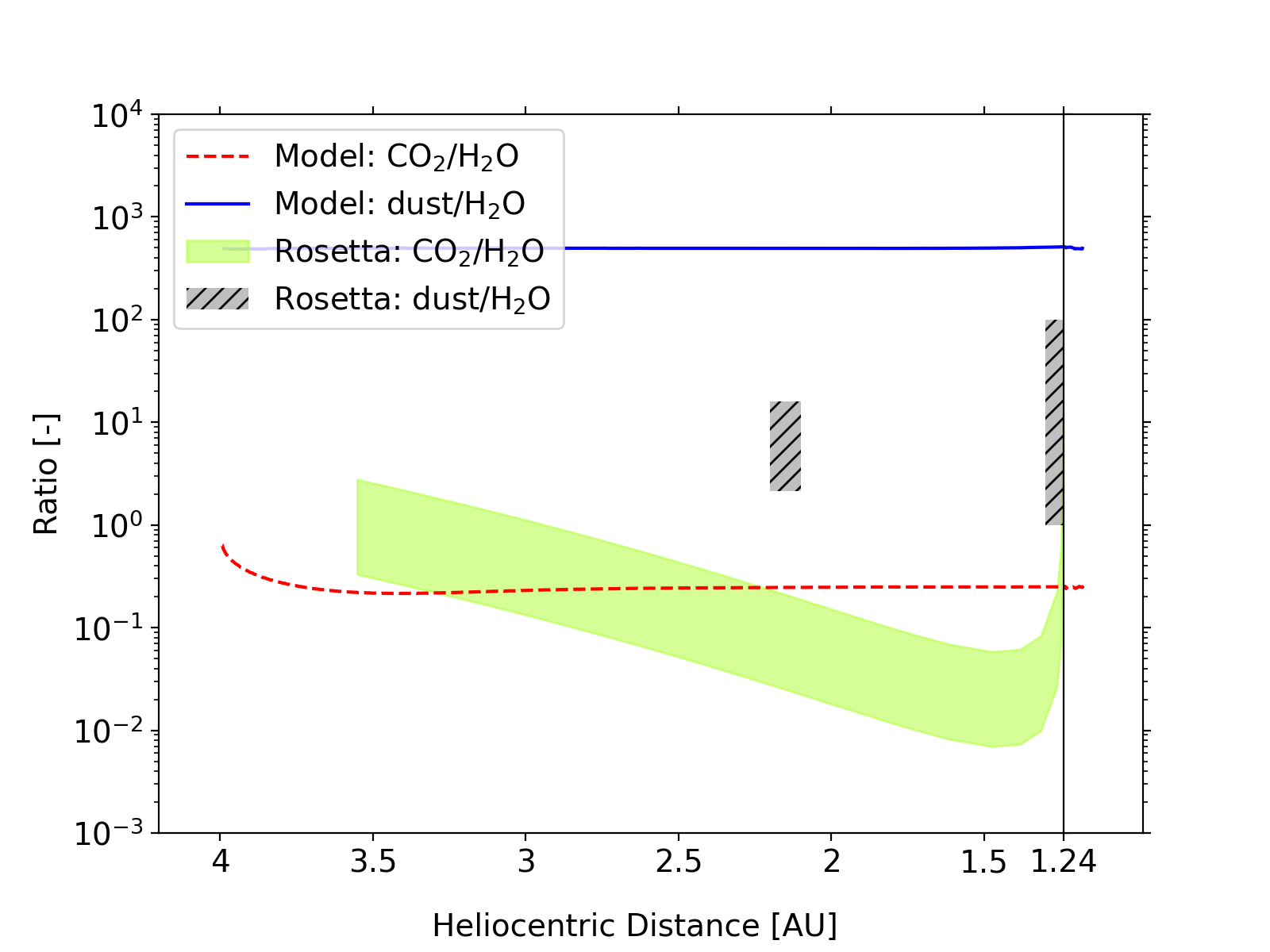}
    \end{minipage}
    \caption{Results for case I.A.i.2a with permanent illumination, namely no day-night-cycle (temperature-dependent thermal conductivity, no scaling, $b=10\,\mathrm{mm}$, 2 H$_2$O-depleted layers). Left: Outgassing- and dust-productions rates for the inbound orbit compared to Rosetta measurements at 67P. Right: Ratio between CO$_2$- and H$_2$O-outgassing rates and ratio between dust-production and H$_2$O-outgassing rates for the inbound orbit, compared to the Rosetta measurements.} 
    \label{fig:results_I_2a_A_nightoff} 
\end{figure*}

\begin{figure*}
    \centering
    \begin{minipage}[b]{.48\linewidth}
    \includegraphics[width=\columnwidth]{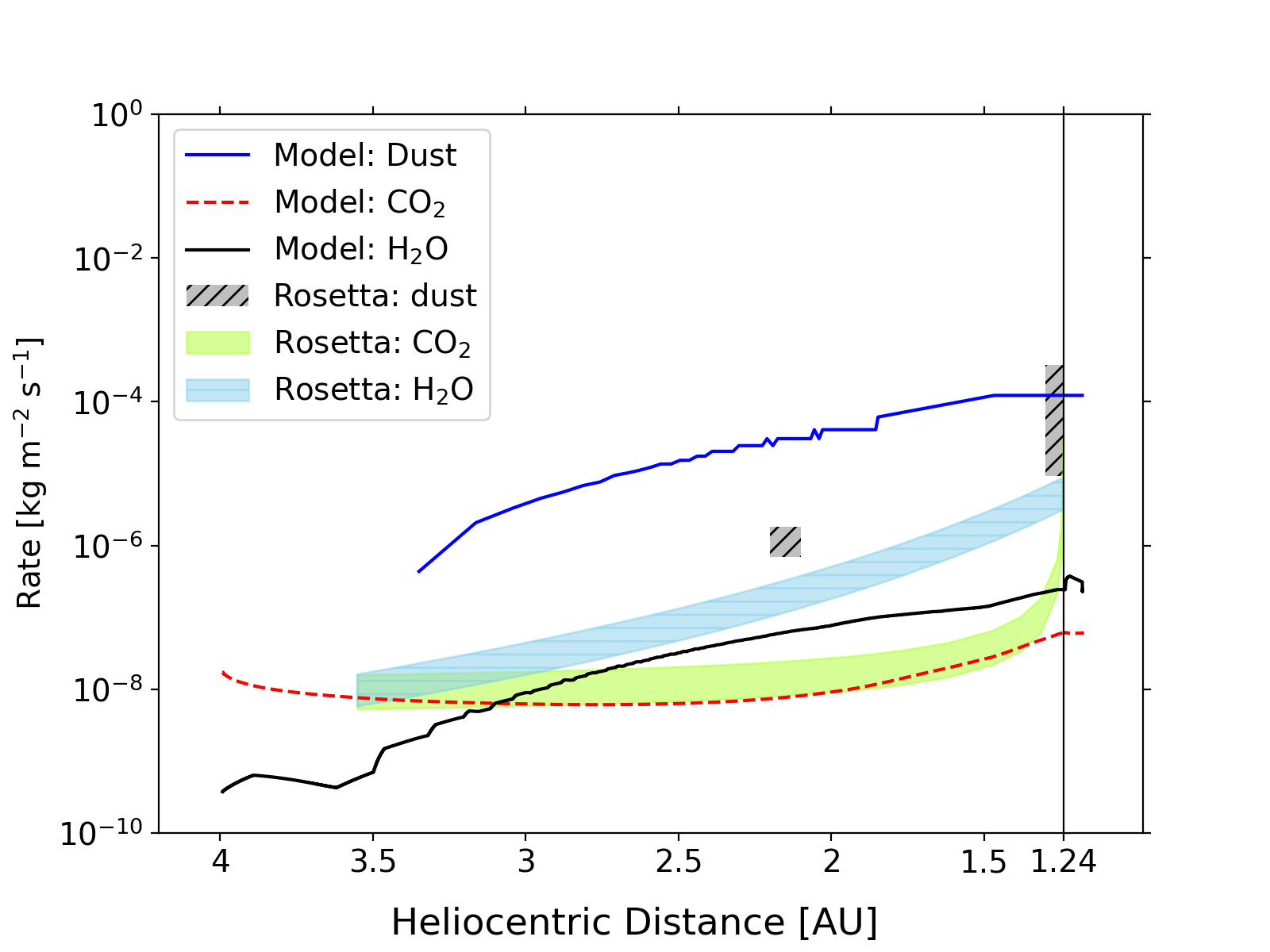}
    \end{minipage}
    \hspace{.03\linewidth}
    \begin{minipage}[b]{.48\linewidth}
    \centering
    \includegraphics[width=\columnwidth]{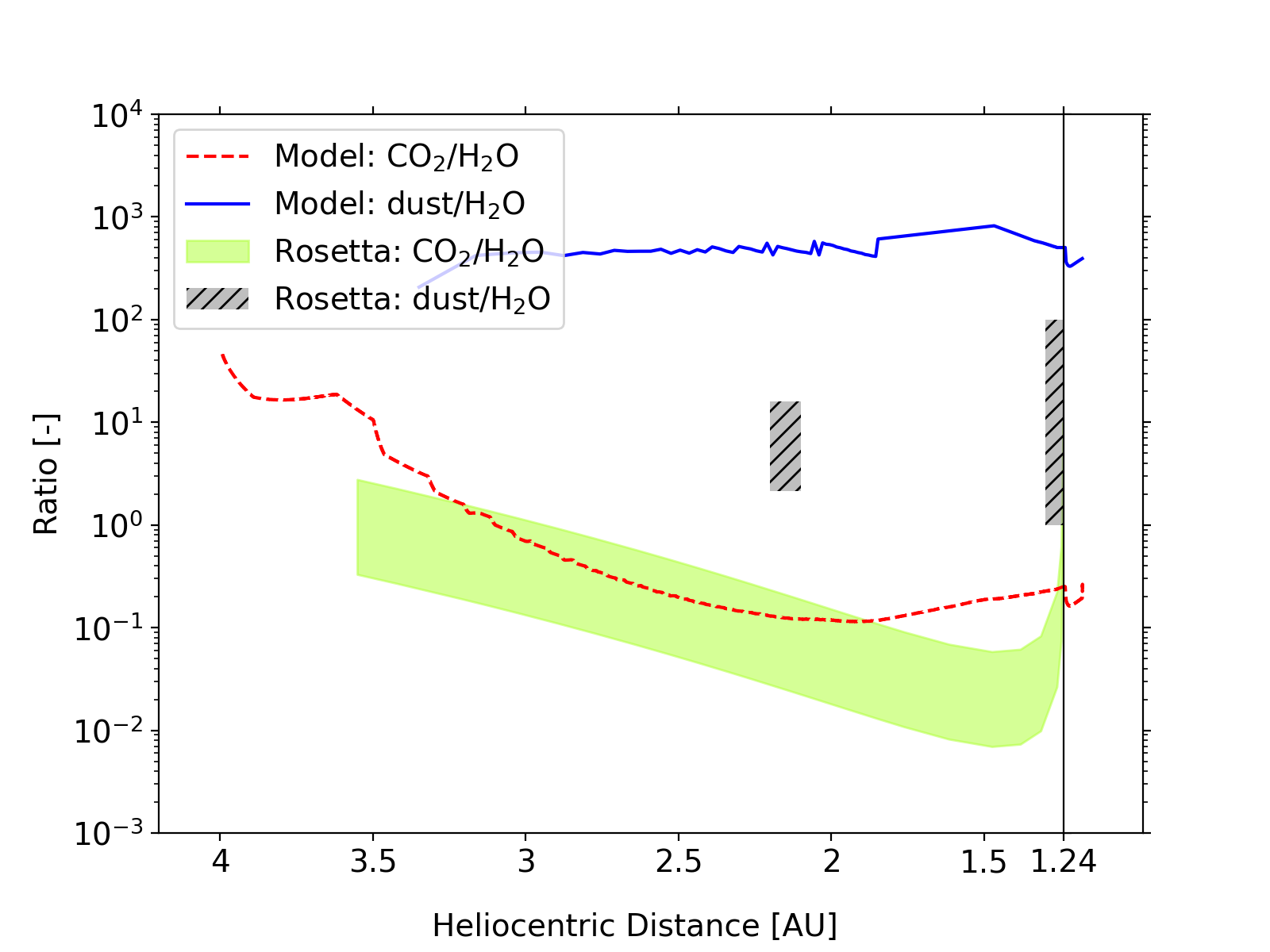}
    \end{minipage}
    \caption{Results for case II.B.i.3a (constant thermal conductivity, $\Phi$ scaling, $b=10\,\mathrm{mm}$, H$_2$O-depleted layers once per day). Left: Outgassing- and dust-productions rates for the inbound orbit compared to Rosetta measurements at 67P. Right: Ratio between CO$_2$- and H$_2$O-outgassing rates and ratio between dust-production and H$_2$O-outgassing rates for the inbound orbit, compared to the Rosetta measurements.} 
    \label{fig:results_II_3a_B} 
\end{figure*}

First, in Section \ref{sec:results_sub:macro} we present the results of the different cases with temperature-dependent thermal conductivity (cases I in Figure \ref{fig:Overview}), followed in Section \ref{sec:results_sub:micro} by the results from the cases with constant thermal conductivity (cases II in Figure \ref{fig:Overview}). The variation of the diffusion length scale (cases ii) can be found in Section \ref{sec:results_sub:low_b}. For each case, we plotted (see Figure \ref{fig:results_I_3a_A} as an example) the modelled outgassing rates and dust production rates (averaged over one comet day) versus heliocentric distance (left-hand side plots), as well as the ratio between the CO$_2$ and H$_2$O outgassing rates and the ratio between the dust production and the H$_2$O outgassing rates (right-hand side plots). In the main part of this paper, we show the best-fitting cases (Figures \ref{fig:results_I_3a_A}, \ref{fig:results_II_3a_B} and \ref{fig:results_I_2a_b_A}) and adapted cases without day-night cycles (Figures \ref{fig:results_I_2a_A_nightoff}, \ref{fig:results_II_2a_A_1E-4_permanent} and \ref{fig:results_II_2a_A_1E-1_permanent}). The plots for all further cases can be found in the Supplementary Material. A summary of our results and an estimation for the agreement to the Rosetta data is presented in Table \ref{Tab:Results_Summary}. A good agreement is marked with ''+", a partly agreement with ''o" and no agreement with ''-". This classification is only considered to be qualitative. For dust production rates, we assign a rate higher than the expected range as a good agreement, because higher values could be scaled down due to varying illumination conditions on the comet.

\begin{table*}
\caption{Summary of the results from our model variations and an estimation for the agreement to the Rosetta data. A good agreement is marked with ''+", a partly agreement with ''o" and no agreement with ''-". This classification is only qualitative. See Fig \ref{fig:Overview} for the corresponding activity mechanism. (*): The size of the ejected dust is fixed to this value due to the activity mechanism.} 
\label{Tab:Results_Summary}

\newcolumntype{C}{>{\centering\arraybackslash}X}
		\begin{tabularx}{\textwidth}{C|C|C|C|CCCCCC}
Thermal conductivity    & Scaling of outgassing rate & Diffusion length  & Activity mechanism  & H$_2$O outgassing rate & CO$_2$ outgassing rate & CO$_2$/H$_2$O ratio & Dust production rate & Dust/H$_2$O ratio & Size of ejected dust \\ \hline
I: & A: No  & i:  & 1a  & - & - & - & no activity & - & -     \\
Temperature &  scaling & $b=10\,\mathrm{mm}$ & 1b  & - & - & - & + & - & $1 \,\mathrm{cm}$  \\
dependent &              &  & 1c  & - & - & - & + & - & $1 \,\mathrm{cm}$   \\
          &              &  & 2a  & - & + & o & + & - & $1 \,\mathrm{cm}$*  \\
          &              &  & 2a, no day/night  & - & o & - & + & - & $1 \,\mathrm{cm}$*  \\
          &              &  & 2b  & - & - & - & + & - & $1 \,\mathrm{cm}$*    \\
          &              &  & 3a  & o & + & o & + & - & $1-1.5 \,\mathrm{cm}$  \\
          &              &  & 3b  & - & - & - & + & - & $1-4.5 \,\mathrm{cm}$ \\ \cline{3-10}
 &                     & ii:  & 1a  & - & - & - & + & - & $1 \,\mathrm{cm}$ \\
 &  & $b=10\,\mathrm{\mu m}$ & 1b  & - & - & - & + & - & $1 \,\mathrm{cm}$  \\
          &              &  & 1c  & - & - & - & + & - & $1 \,\mathrm{cm}$   \\
          &              &  & 2a  & - & + & o & + & - & $1 \,\mathrm{cm}$*  \\
          &              &  & 2b  & - & - & - & + & - & $1 \,\mathrm{cm}$*    \\ \cline{2-10}
          & B: Volume    & i: & 1a  & - & - & -  & no activity & - & - \\
          & filling      & $b=10\,\mathrm{mm}$ & 1b  & - & - & - & + & - & $1 \,\mathrm{cm}$  \\
          & factor       &  & 1c  & - & - & - & + & - & $1 \,\mathrm{cm}$   \\
          &              &  & 2a  & - & + & o & + & - & $1 \,\mathrm{cm}$*  \\
          &              &  & 2b  & - & - & - & + & - & $1 \,\mathrm{cm}$*    \\
          &              &  & 3a  & o & + & o & + & - & $1-1.5 \,\mathrm{cm}$  \\
          &              &  & 3b  & - & - & - & + & - & $1-4.5 \,\mathrm{cm}$ \\ \hline
II: & A: No & i:  & 1a & - & - & - & no activity & - & - \\
Constant  &    scaling  &  $b=10\,\mathrm{mm}$ & 1b & - & - & - & + & - & $1 - 1.5 \,\mathrm{cm}$ \\
$10^{-2}$   &              &  & 1c & - & - & - & + & - & $1 \,\mathrm{cm}$ \\
$\mathrm{W/(K \,m)}$ &              &  & 2a & - & + & o & + & - & $1 \,\mathrm{cm}$*  \\
           &              &  & 2b & - & - & - & + & - & $1 \,\mathrm{cm}$*  \\
           &              &  & 3a & o & + & o & + & - & $1 \,\mathrm{cm}$ \\
           &              &  & 3b & - & - & - & + & - & $1 - 5 \,\mathrm{cm}$ \\  \cline{2-10}
 & B: Volume    & i:  & 1a & - & - & - & no activity & - & - \\
 & filling      & $b=10\,\mathrm{mm}$ & 1b & - & - & - & + & - & $1 \,\mathrm{cm}$ \\
           &  factor      &  & 1c & - & - & - & + & - & $1 \,\mathrm{cm}$ \\
           &              &  & 2a & - & + & o & + & - & $1 \,\mathrm{cm}$*  \\
           &              &  & 2b & - & - & - & + & - & $1 \,\mathrm{cm}$*  \\
           &              &  & 3a & o & + & o & + & - & $1 \,\mathrm{cm}$ \\
           &              &  & 3b & - & - & - & + & - & $1 - 5 \,\mathrm{cm}$ \\ \cdashline{1-10}
$10^{-4}$ $\mathrm{W/(K \,m)}$ & A: No scaling & i: $b=10\,\mathrm{mm}$ & 2a, no day/night & - & - & o & o & - & $1 \,\mathrm{cm}$*  \\
$10^{-1}$ $\mathrm{W/(K \,m)}$ &  &  & 2a, no day/night & - & o & - & + & - & $1 \,\mathrm{cm}$* \\
           \hline

\end{tabularx}
\end{table*}

\subsection{Temperature-dependent thermal conductivity}
\label{sec:results_sub:macro}
Here, we describe the results for the temperature-dependent heat-transport model (cases I in Figure \ref{fig:Overview}). When assuming an activity mechanism based on the vapour pressure (scenario 1 in Figure \ref{fig:Overview}), the scaling of the outgassing rate (comparison of scenarios A and B in Figure \ref{fig:Overview}) is of minor importance. In both cases, a tensile-strength threshold of $10^{-1} \,\mathrm{Pa}$ (case 1a) leads to no dust activity, due to too low pressures in both cases (see Figures A1 and A3 in Supplementary Materials). However, $10^{-2} \,\mathrm{Pa}$ (case 1b) are reached in both cases by CO$_2$ gas pressure in most parts of the orbit (see Figures A4 and A6 in Supplementary Materials). For all heliocentric distances, the outgassing rate of CO$_2$ exceeds that of H$_2$O, which is in contradiction to the observations. The slope of both rates is shallower than observed and also the absolute values of the outgassing rates lie below the Rosetta data. The dust production is several orders of magnitude too high and the dust-to-gas ratio even exceeds $10^4$ for most of the modelled orbit part, whereas the Rosetta data only reaches up to 100 at perihelion with a high uncertainty. Reducing the threshold pressure to $10^{-3} \,\mathrm{Pa}$ (case 1c)  does not influence the outcome (see Figure A7 and A9 in Supplementary Materials). Therefore, both cases 1b and 1c show pressure-driven dust activity, but do not match the observations. 

For the H$_2$O-depletion-driven activity mechanism (case 2a in Figure \ref{fig:Overview}), where two layers get ejected when they are depleted in H$_2$O, the absolute values of the outgassing rates of CO$_2$ matches the Rosetta data quite well, but the outgassing values for H$_2$O are too low (see Figures A10 and A11 in Supplementary Materials). The slopes do not increase as much as observed near perihelion, but the ratio of these rates is in general agreement with the observations. The dust production rates exceed the measured data in absolute values as well as relative to H$_2$O and the slope is too shallow. However, for perihelion the absolute value is around the upper limit of the observations. 

For the CO$_2$-depletion-driven activity mechanism (scenario 2b in Figure \ref{fig:Overview}), the results drastically change and are similar to the results for pressure-driven activity (see Figures A12 and A13 in Supplementary Materials). The CO$_2$ outgassing rate exceeds that of H$_2$O and shows a shallow slope as well as a high dust-production rate. 

The third activity mechanism, where once per day all depleted layers of H$_2$O (scenario 3a in Figure \ref{fig:Overview}) or CO$_2$ (scenario 3b) are ejected, produces similar results to scenario 2a and 2b. For the H$_2$O-depletion-driven activity (case I.A.i.3a, see Figure \ref{fig:results_I_3a_A}), the absolute values of the CO$_2$ outgassing rates as well as the ratio of the outgassing rates are in the expected order of magnitude. However, also here H$_2$O outgassing is below the observations. The dust production rate is too shallow, but is for perihelion within the upper limit of the observations. Analogously, the mechanism based on CO$_2$ depletion (case 3b) does not match observations at all, as the CO$_2$ outgassing rate exceeds that of H$_2$O. However, it should be noted here that this scenario is the only one resulting in a change of the size of the emitted dust particles, which increases when approaching the Sun up to $4.5 \,\mathrm{cm}$ around perihelion.

For all cases in which activity occurred, the ratio of dust production to H$_2$O production exceeds the expected range by at least an order of magnitude. 

Because of the obliquity of the comet, seasonal effects influence the illumination conditions and, thus, outgassing and dust production rates. To investigate whether this could explain our mismatch, we ran a case with permanent illumination for the whole orbit, namely turning off the day-night cycles. This is of course only realistic for the polar-day phases, but the most extreme scenario with the highest energy input. If this does not result in high enough H$_2$O outgassing rates, the seasonal effect cannot explain the mismatch of our model. The results of this simulation are shown in Figure \ref{fig:results_I_2a_A_nightoff}. We assume here the second activity mechanism driven by H$_2$O depletion, because as shown in the other cases, activity cannot be explained by pressure thresholds. In this case, all rates show a similar slope, and the ratios are therefore nearly constant for this part of the orbit. The H$_2$O outgassing rates exceed the expected range farther out than roughly $2.5 \,\mathrm{au}$ but remain lower than expected near perihelion. The constant slopes do not match the observations. Therefore, we can conclude that including the seasonal effects in our model cannot result in matching H$_2$O outgassing rates around perihelion. 

\subsection{Temperature-independent thermal conductivity}
\label{sec:results_sub:micro}
In general, the results with constant thermal conductivity (cases II in Figure \ref{fig:Overview}) do not differ significantly from the cases with temperature dependency. The pressure-driven activity mechanism with no scaling of the outgassing rate for a threshold pressure of $10^{-2} \,\mathrm{Pa}$ (case 1b) results in activity (see Figures B3 and B4 in Supplementary Materials). As for the temperature-dependent cases, this threshold value is reached by CO$_2$ from the beginning of the simulation and results in outgassing and dust production rates not matching the observation. CO$_2$ outgassing exceeds H$_2$O by a factor of $\geq 8$ and even exceeds the scale of the plot, which is $10^4$, in several cases. Figure \ref{fig:results_II_3a_B} shows the results of the activity mechanism with ejection once per day driven by H$_2$O depletion, which matches best the observations, but still without matching the H$_2$O outgassing rates and the ratio of dust production to H$_2$O production.

In conclusion, the temperature dependency of the thermal conductivity seems to not influence our results significantly. Therefore, our model does not allow for a conclusion regarding this property.

\begin{figure*}
    \centering
    \begin{minipage}[b]{.48\linewidth}
    \includegraphics[width=\columnwidth]{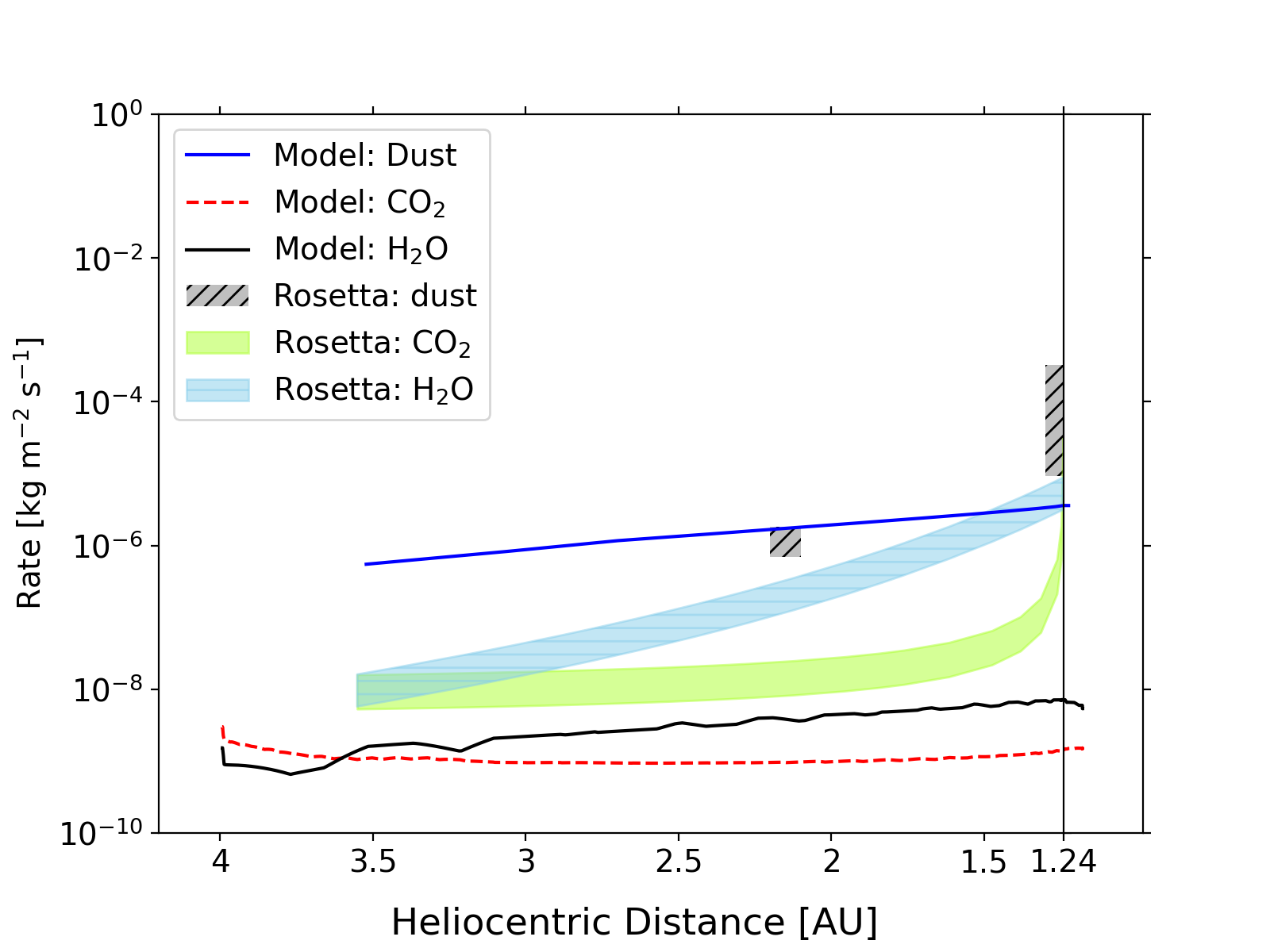}
    \end{minipage}
    \hspace{.03\linewidth}
    \begin{minipage}[b]{.48\linewidth}
    \centering
    \includegraphics[width=\columnwidth]{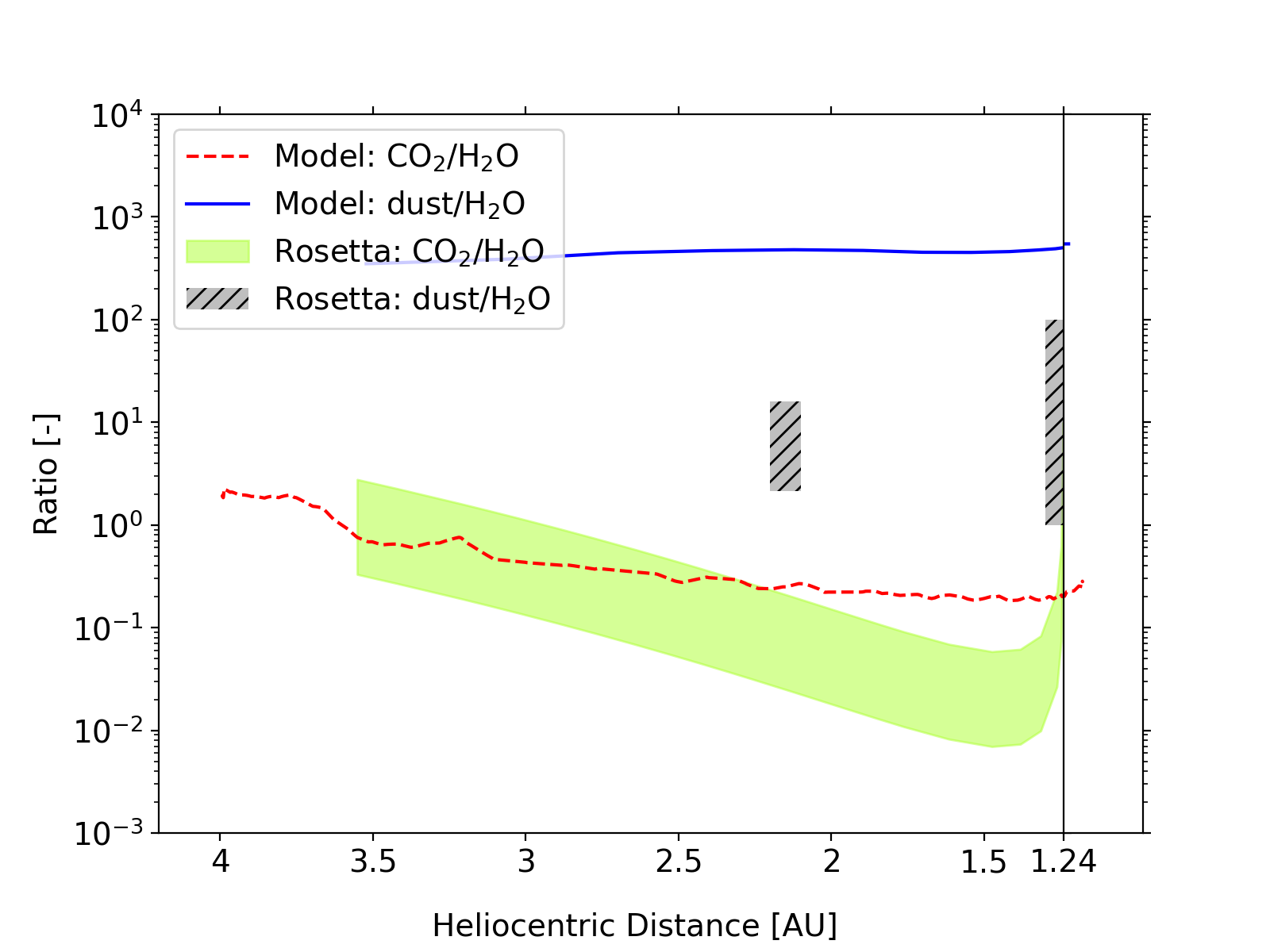}
    \end{minipage}
    \caption{Results for an adaptation of case II.A.i.2a but with a constant thermal conductivity of $10^{-4}\,\mathrm{W/(K \,m)}$ with permanent illumination, namely no day-night-cycle (constant thermal conductivity, $\Phi$ scaling, $b=10\,\mathrm{mm}$, 2 H$_2$O-depleted layers). Left: Outgassing- and dust-productions rates for the inbound orbit compared to Rosetta measurements at 67P. Right: Ratio between CO$_2$- and H$_2$O-outgassing rates and ratio between dust-production and H$_2$O-outgassing rates for the inbound orbit, compared to the Rosetta measurements.} 
    \label{fig:results_II_2a_A_1E-4_permanent} 
\end{figure*}

\begin{figure*}
    \centering
    \begin{minipage}[b]{.48\linewidth}
    \includegraphics[width=\columnwidth]{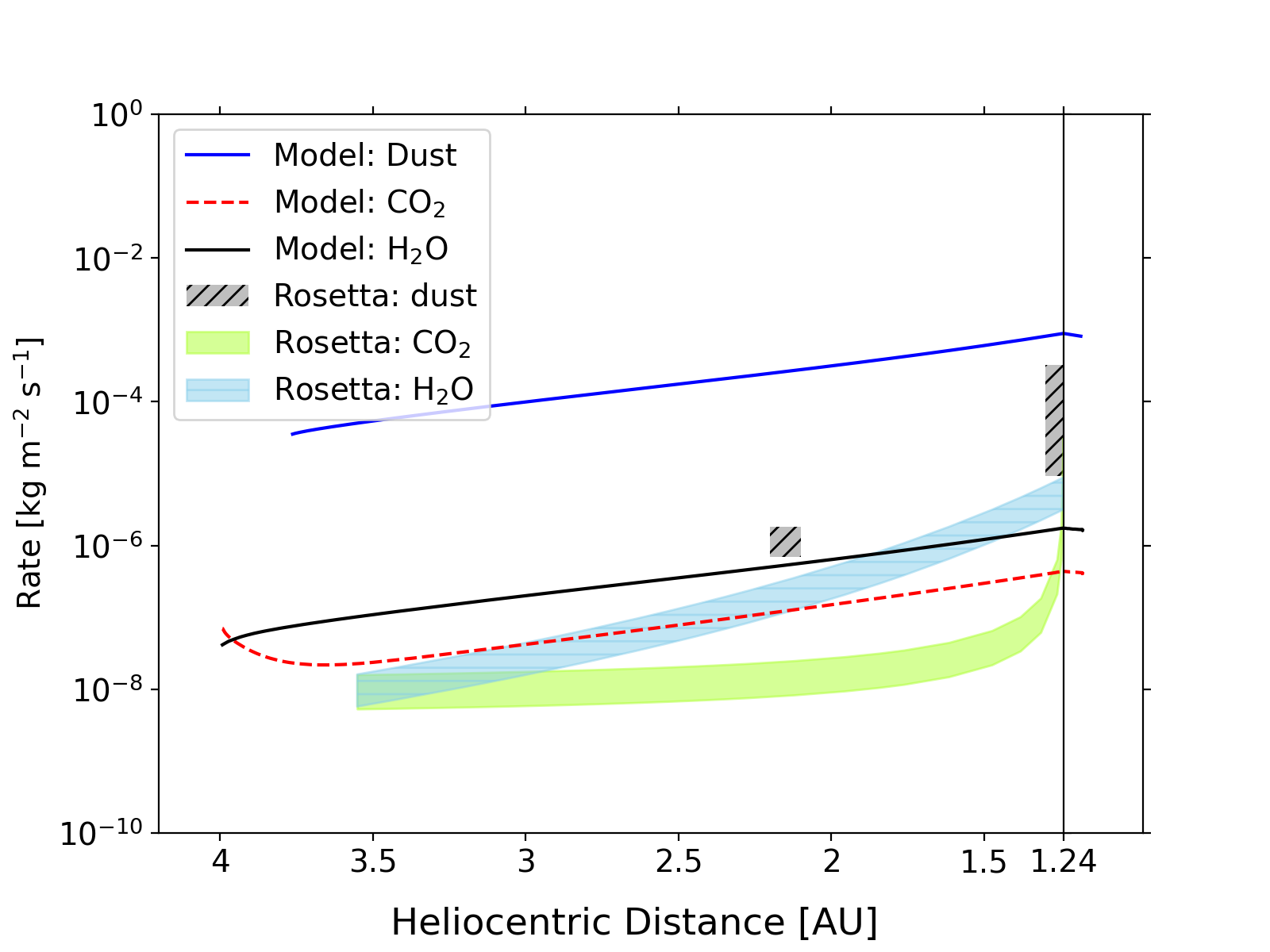}
    \end{minipage}
    \hspace{.03\linewidth}
    \begin{minipage}[b]{.48\linewidth}
    \centering
    \includegraphics[width=\columnwidth]{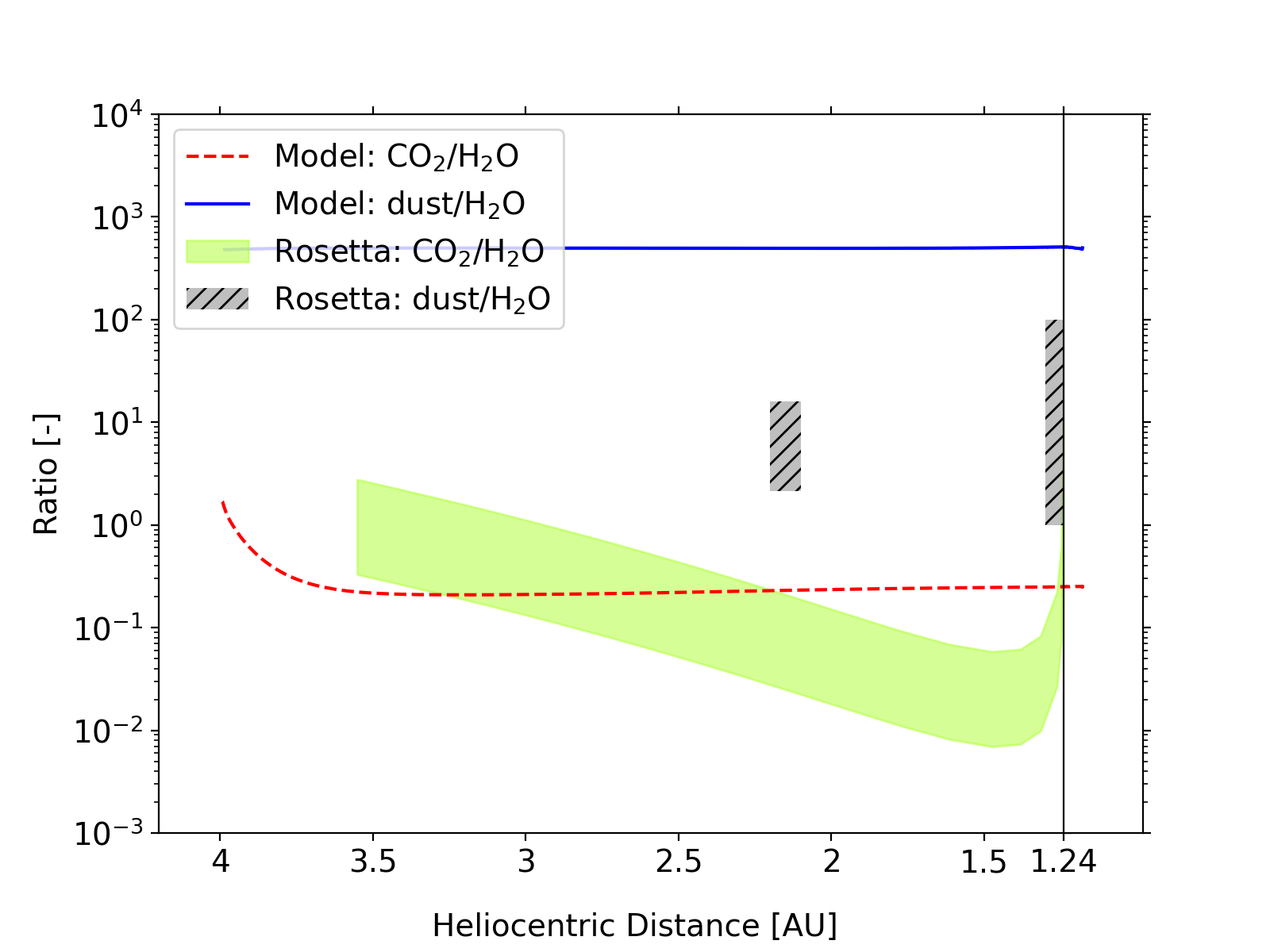}
    \end{minipage}
    \caption{Results for an adaptation of case II.A.i.2a but with a constant thermal conductivity of $10^{-1}\,\mathrm{W/(K \,m)}$ with permanent illumination, namely no day-night-cycle (constant thermal conductivity, $\Phi$ scaling, $b=10\,\mathrm{mm}$, 2 H$_2$O-depleted layers). Left: Outgassing- and dust-productions rates for the inbound orbit compared to Rosetta measurements at 67P. Right: Ratio between CO$_2$- and H$_2$O-outgassing rates and ratio between dust-production and H$_2$O-outgassing rates for the inbound orbit, compared to the Rosetta measurements.} 
    \label{fig:results_II_2a_A_1E-1_permanent} 
\end{figure*}

To investigate whether a case with a different constant thermal conductivity could better match the observations, we performed two additional runs using the maximum energy input case of a permanent illumination without day-night cycles. We used a constant thermal conductivity of $10^{-4}\,\mathrm{W/(K \,m)}$ and $10^{-1}\,\mathrm{W/(K \,m)}$ and the results are presented in Figure \ref{fig:results_II_2a_A_1E-4_permanent} and \ref{fig:results_II_2a_A_1E-1_permanent}, respectively. A very low thermal conductivity leads to very low outgassing rates and also reduces the dust production rate, not reproducing the observations. With higher thermal conductivity, higher outgassing and dust production rates are reached, exceeding observations for most parts of the orbit, but for perihelion, the water production is still a factor of roughly 10 below expectations. 


\subsection{Reduced diffusivity}
\label{sec:results_sub:low_b}

As described in Section \ref{sec:TPM:ScalingOutgassing}, we additionally investigated the influence of a reduced diffusivity (cases ii in Figure \ref{fig:Overview}). If the void space between the pebbles is filled with fluffy dust, as observed at comet 67P \citep{Fulle.2017}, the flow for molecules into space would be hindered. This can be described by a reduced $b$ parameter, which we chose as $b = 10\,\mathrm{\mu m} = 10^{-3}$ pebble diameter (scenario b). To save computation time, we did not run all models with this variant, but selected five. As the influence of the scaling of the outgassing rate as well as of the temperature dependency of the thermal conductivity, we chose one of each scenario, namely the case with no scaling (A) and temperature-dependent conductivity (I). The difference between the two artificial activity mechanism is also small. Because an ejection once per day (scenarios 3a, 3b) lacks any physical explanation, we decided to run models with the layer-depletion mechanism (scenarios 2a, 2b). Because a higher pressure is expected, we ran the model for all pressure-driven activity mechanisms (scenarios 1a, 1b, 1c). 

The results of the best-fitting case, namely the artifical H$_2$O-depletion-driven activity mechanism, is shown in Figure \ref{fig:results_I_2a_b_A}, whereas the others are given in the Supplementary Materials. With the reduced $b$ parameter, a pressure of $0.1 \,\mathrm{Pa}$ can be reached and activity occurs, but it is always driven by CO$_2$, resulting in non-matching outgassing ratios between CO$_2$ and H$_2$O. Therefore, also in these cases a pressure-dependent or an artificially CO$_2$-depletion-driven activity mechanism does not match the observations. In the case of the artificially H$_2$O-depletion-driven activity, the outgassing rate of H$_2$O lies clearly below the expectation, but it shows a comparable slope, which was not reached in cases with higher diffusion length. The CO$_2$ outgassing matches the lower bound of the observations quite well. Also, the dust production fits quantitatively well to the two data points, however, due to the low H$_2$O production rate, the dust-to-gas ratio is too high, as well as the CO$_2$-to-H$_2$O ratio. In conclusion for this scenario, a reduced diffusion length reduces the absolute H$_2$O-outgassing rates, but increases its slope towards the observed slope. 

\begin{figure*}
    \centering
    \begin{minipage}[b]{.48\linewidth}
    \includegraphics[width=\columnwidth]{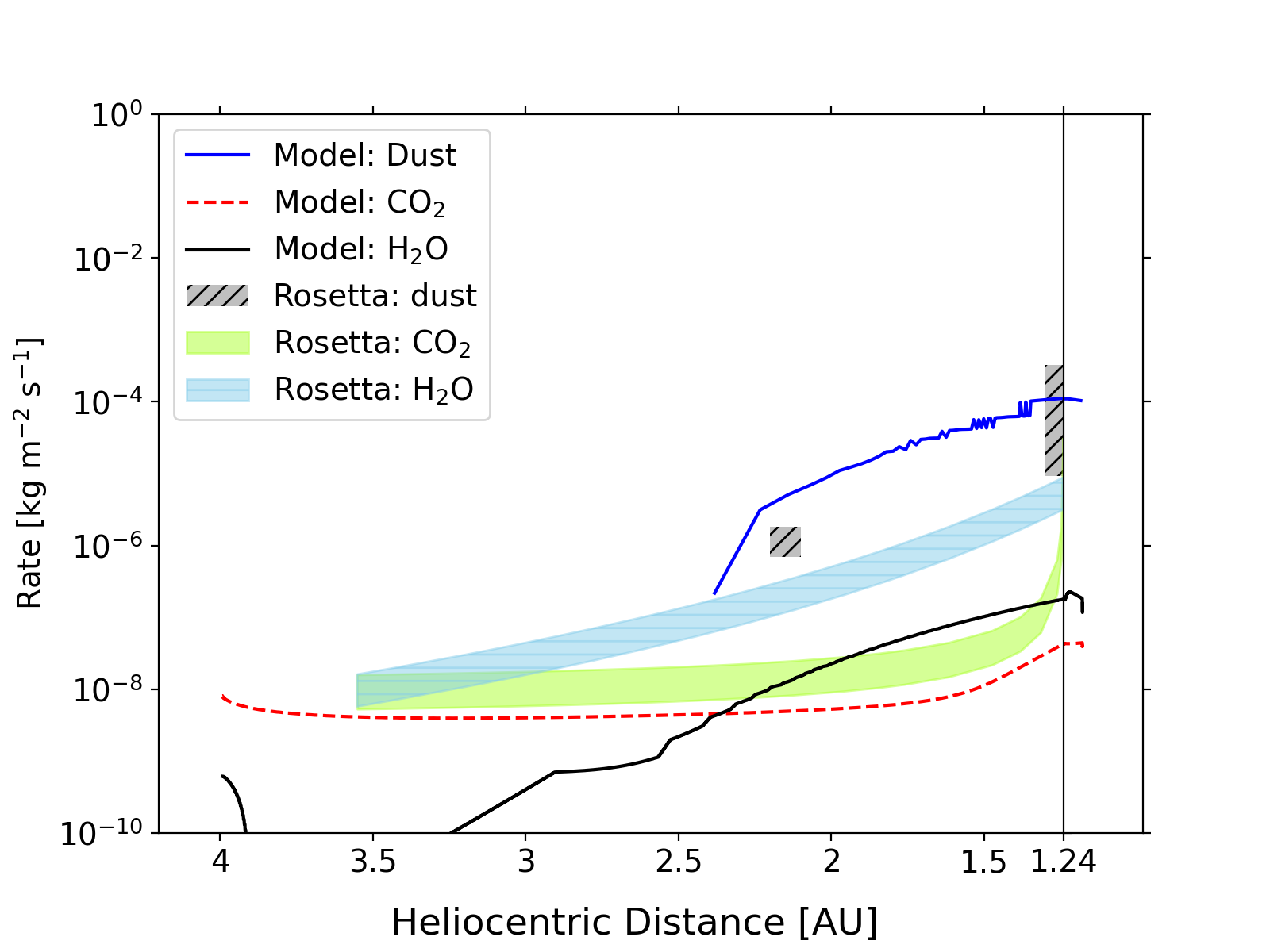}
    \end{minipage}
    \hspace{.03\linewidth}
    \begin{minipage}[b]{.48\linewidth}
    \centering
    \includegraphics[width=\columnwidth]{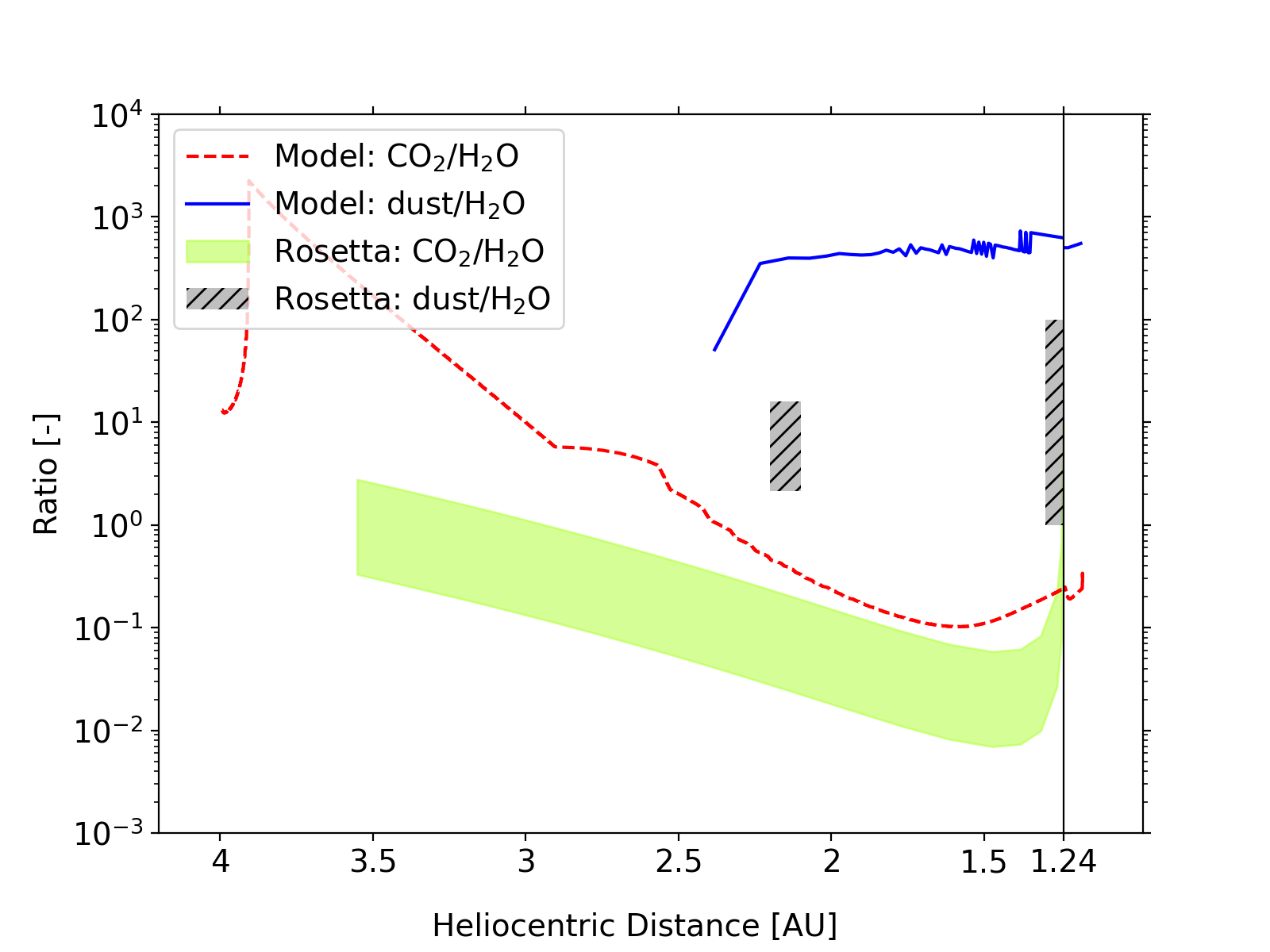}
    \end{minipage}
    \caption{Results for case I.A.ii.2a (temperature-dependent thermal conductivity, no scaling, $b=10\,\mathrm{\mu m}$, 2 H$_2$O-depleted layers). Left: Outgassing- and dust-productions rates for the inbound orbit compared to Rosetta measurements at 67P. Right: Ratio between CO$_2$- and H$_2$O-outgassing rates and ratio between dust-production and H$_2$O-outgassing rates for the inbound orbit, compared to the Rosetta measurements.} 
    \label{fig:results_I_2a_b_A} 
\end{figure*}

\section{How to overcome tensile strength?}
\label{sec:results:pressure}

As can be seen in the previous Section, our model does not reach pressures of 0.1 Pa or higher with an expected diffusivity of a pebble structure with large voids, but only with a reduced diffusivity. In this Section, we intend to investigate which pressures can be reached theoretically and at which distances from the Sun. Therefore, we used several approaches to constrain maximum pressures of H$_2$O, CO$_2$ and CO. 
First, we calculated the maximum surface temperature and compared the result with a minimal temperature from sublimation. These heliocentric distance estimations can be compared to observed activities of (far) comets. 

The maximum surface temperature $T_{\mathrm{max}}$ corresponds to the case in which all absorbed solar energy is re-emitted by thermal radiation. This temperature reads
\begin{equation}
    T_{\mathrm{max}} = \left(\frac{(1-A) \, I_0}{\sigma \, \epsilon \, r^2} \right)^{1/4},
    \label{eq:max_temperature}
\end{equation}
with the albedo $A$, the solar constant $I_0$, the heliocentric distance $r$ in au, the Stefan-Boltzmann constant $\sigma$ and the emissivity $\epsilon$. Here, we assume an albedo of $A=0.055$. 

The opposite case is that all solar energy is consumed by sublimation of one specific ice $X$. The absorbed solar energy divided by the latent heat $\Lambda_X(T)$ of the regarding ice $X$ results in the maximum amount of ice mass that can be sublimated per unit time and area. Combined with the Hertz-Knudsen equation, the corresponding temperature can be calculated, which is the minimum temperature $T_{\mathrm{min,X}}$. It can be determined by 
\begin{equation}
    \frac{(1-A) \, I_0}{r^2 \Lambda_X(T)} = p_{\mathrm{sat},X}(T) \sqrt{\frac{m_X}{2 \pi k T}}, 
    \label{eq:temp_sublimation}
\end{equation}
with the saturation pressure $ p_{\mathrm{sat},X}(T)$, the mass of the regarding molecule $m_X$ and the Boltzmann constant $k$. It is also possible to directly calculate a resulting pressure $p_X$ from the balance of absorbed solar energy and latent heat through
\begin{equation}
  p_X =   \frac{(1-A) \, I_0}{r^2 \Lambda_X(T)}  \sqrt{\frac{2 \pi k T}{m_X}}.
    \label{eq:temp_sublimation_pressure}
\end{equation}
Both equations \ref{eq:temp_sublimation} and \ref{eq:temp_sublimation_pressure} are not analytical solvable for $T$ and they are not unambiguous. For the first one, the assumption of a saturation pressure might not be right. For the second one, a temperature $T$ inside the velocity term needs to be assumed. However, its influence is small. We numerically found the solutions for the minimal temperatures $T_{\mathrm{min},X}$ in Equation \ref{eq:temp_sublimation} for water ice, carbon dioxide ice and carbon monoxide ice and used these temperatures in Equation \ref{eq:temp_sublimation_pressure}. As these equations are not unambiguously solvable, we used this simplification, although it is physically not correct. Therefore, the pressure used in Equation \ref{eq:temp_sublimation} is not exactly equal to the pressure resulting from Equation \ref{eq:temp_sublimation_pressure}, but the difference is not significant. The relations for $\Lambda_X(T)$ and $p_{\mathrm{sat},X}(T)$ as well as the values used here can be found in Appendix \ref{App:TPM}.

\begin{figure}
    \centering
    \includegraphics[width=\columnwidth]{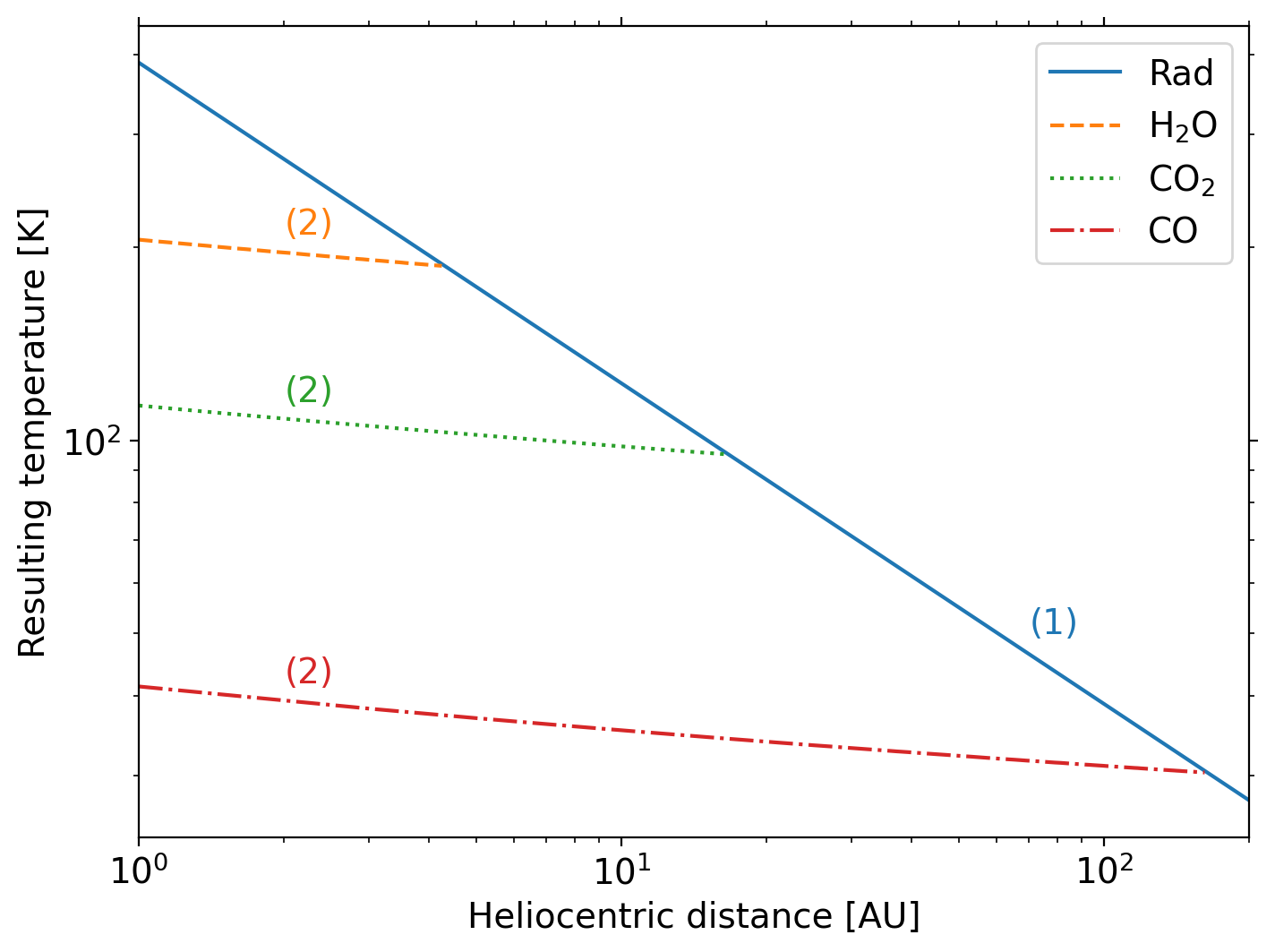}
    \caption{Maximum possible temperature assuming all energy is re-radiated (Equation \ref{eq:max_temperature}, blue, solid, labelled with 1) and minimum temperature assuming all energy is consumed by sublimation (Equation \ref{eq:temp_sublimation}, labelled with 2) depending on heliocentric distance for water (orange, dashed), carbon dioxide (green, dotted) and carbon monoxide ice (red, dashed-dotted). The intersection corresponds to the maximum activity distance, which is at $4.5 \,\mathrm{au}$ for H$_2$O, $13.7 \,\mathrm{au}$ for CO$_2$ and $194 \,\mathrm{au}$ for CO.}
    \label{fig:Maximal_Actvity_Distance}
\end{figure}

The results of these calculations for H$_2$O, CO$_2$ and CO are plotted in Figure \ref{fig:Maximal_Actvity_Distance}. The intersection of the maximum temperature by radiation and the minimum temperature by sublimation refers to the maximum activity distance from the Sun for the respective ices, assuming that the activity is a process in balance produced by the solar energy. Dust activity driven by sublimation of water ice is only possible within $4.5 \,\mathrm{au}$, by sublimation of CO$_2$ within $13.7 \,\mathrm{au}$ and for CO within $194 \,\mathrm{au}$. This is consistent with observations of activity in the Solar System and with the activity model proposed by \citet{Fulle.2022b}. For H$_2$O ($3.8 \,\mathrm{au}$) and CO$_2$ ($13 \,\mathrm{au}$), the onset heliocentric distance of \citet{Fulle.2022b} lie close to the distances calculated here. It should be noted that the ‘’minimum temperature´´ is not the minimum temperature that a body at this distance has, especially in the interior, where it can be much cooler. 

\begin{figure}
    \centering
    \includegraphics[width=\columnwidth]{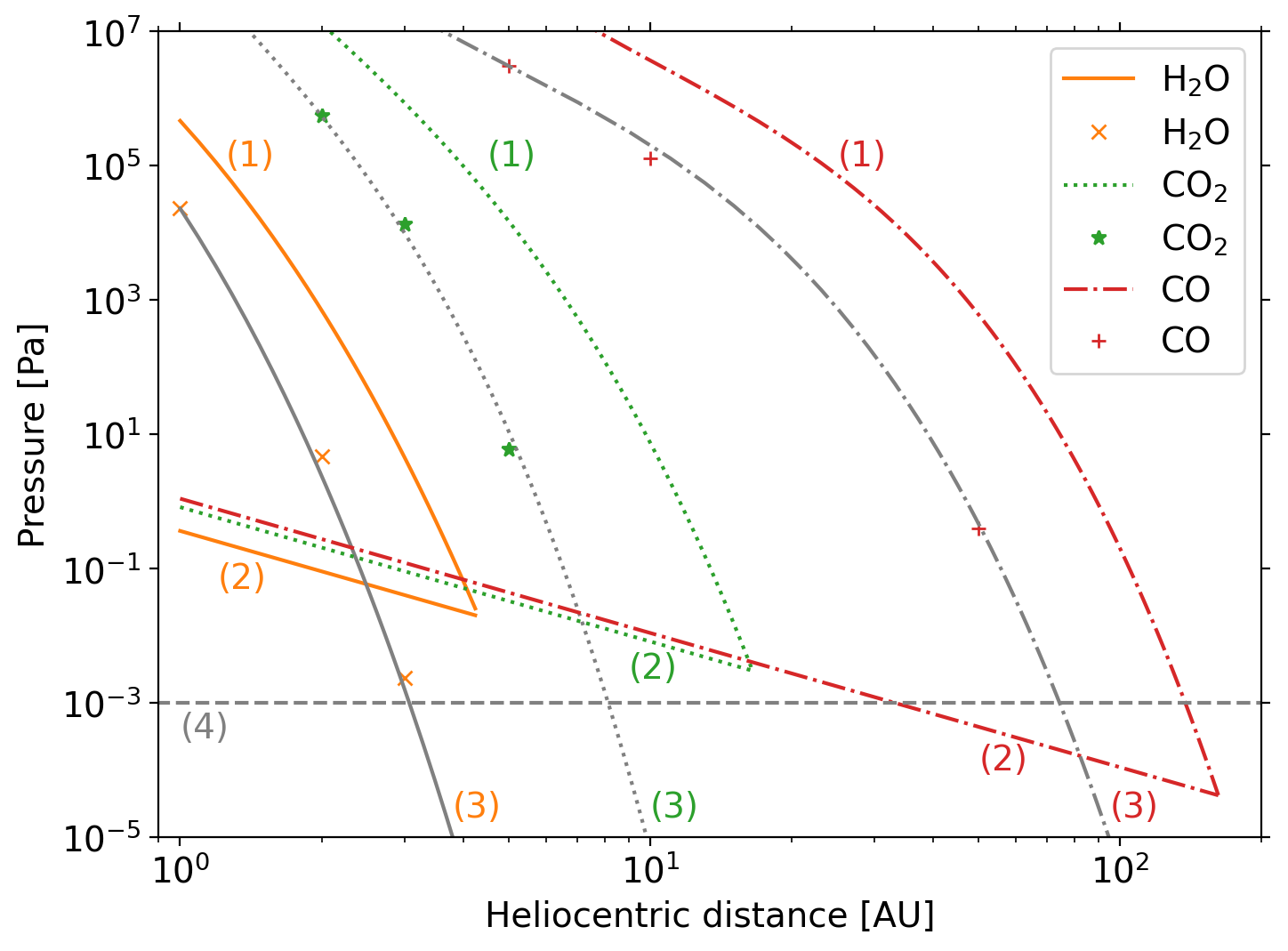}
    \caption{Saturation pressure of water (orange, solid), carbon dioxide (green, dotted) and carbon monoxide ice (red, dashed-dotted), depending on the heliocentric distance corresponding to the minimal and maximal temperatures in Figure  \ref{fig:Maximal_Actvity_Distance}. Lines labelled with (1) correspond to the cases in which all energy is re-radiated. lines labelled with (2) correspond to the cases in which all energy is consumed by sublimation of the respective ice. The symbols indicate the pressures at a depth of $1 \,\mathrm{cm}$ resulting from our adapted model without outgassing into space or deeper layers. The grey lines labelled with (3) are fit functions to that data, which can be found in Appendix \ref{App:fits}. The horizontal grey dashed line labelled with (4) indicates the pressure resulting from the weight of a $1 \,\mathrm{cm}$ thick dust layer with a density of $532 \,\mathrm{kg/m^3}$ and an acceleration of gravity of $2 \times 10^{-4} \,\mathrm{m/s}$.}
    \label{fig:Maximal_Actvity_Distance_pressure}
\end{figure}

Regarding the maximum reachable pressures, we calculated the saturation pressure with Equation \ref{eq:temp_sublimation_pressure}, corresponding to the shown temperatures presented in Figure \ref{fig:Maximal_Actvity_Distance_pressure}, resulting in a large range of possible maximum pressures. The lines end when the maximum activity distance is reached. The horizontal line indicates the pressure resulting from the mass of a $1 \,\mathrm{cm}$ dust layer with a density of $532 \,\mathrm{kg/m^3}$ and an acceleration of gravity of $2 \times 10^{-4} \,\mathrm{m/s}$, which is $10^{-3} \,\mathrm{Pa}$. 

\begin{figure}
    \centering
    \includegraphics[width=\columnwidth]{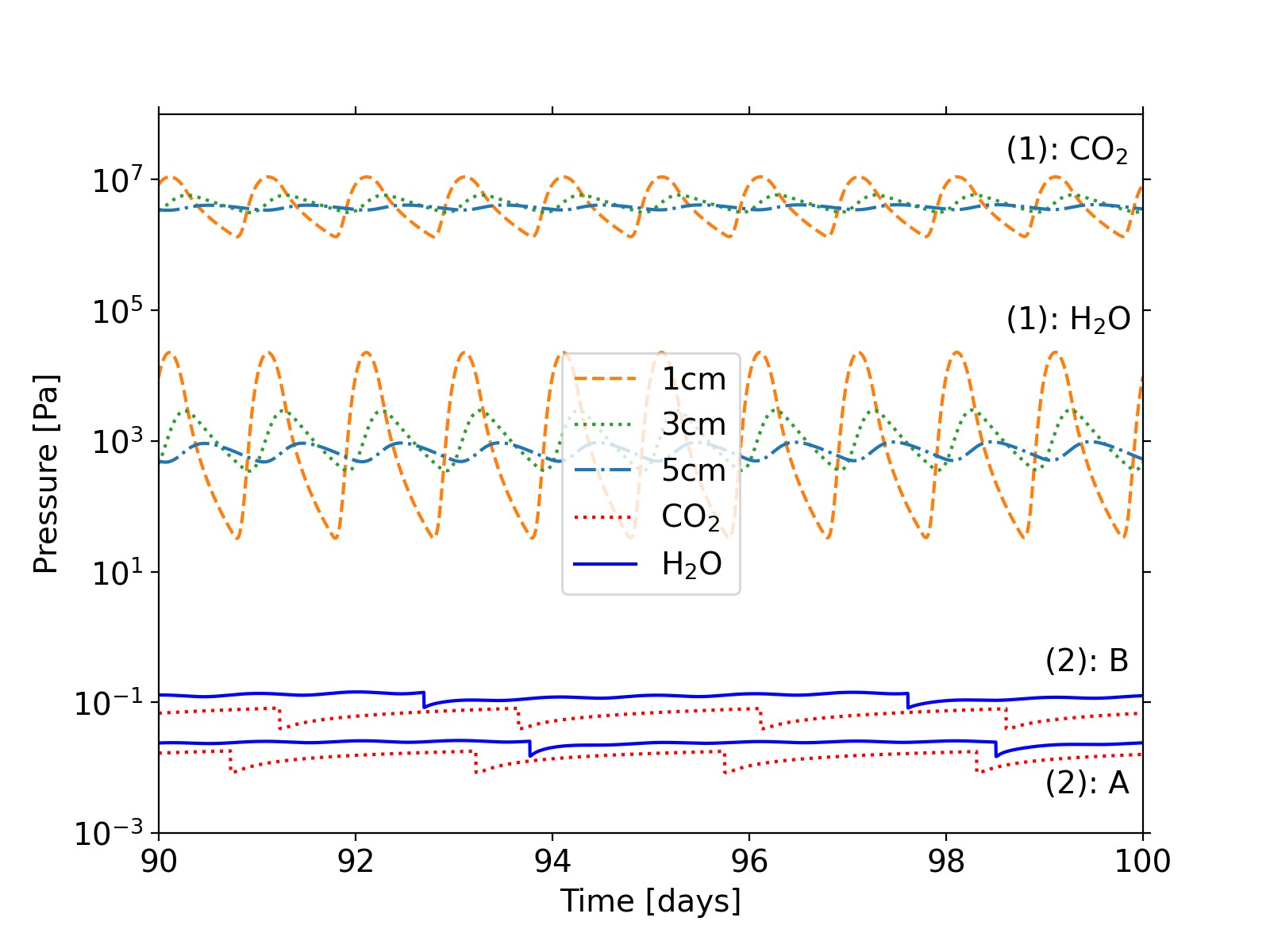}
    \caption{Maximum pressure of different models presented in Sections \ref{sec:results:pressure} (1, no gas transport between layers) and \ref{sec:TPM} (2, standard model with $b= 10\,\mathrm{mm}$ for the case of A: no scaling and B: $\Phi$ scaling) at a distance of $1 \,\mathrm{au}$ after 90 simulated days without dust ejection. For model (1) the maximum pressure in depth of $1$, $3$, and $5\,\mathrm{cm}$ is plotted. One scenario simulated only H$_2$O ice with 3:1 dust-to-ice mass ratio and the other simulated only CO$_2$ ice with 3:1 dust-to-ice mass ratio. For model (2) only the total maximum pressure is given, because the sublimation front recedes with time.}
    \label{fig:pressures_tpm}
\end{figure}

However, the upper limit is not physically possible, because energy will always be consumed by sublimation and heat conduction into the interior. To estimate more reasonable maximum values, we used our thermophysical model (temperature-dependent thermal conductivity, no scaling of outgassing rate, no activity) and adapted it, by reducing the diffusivity to zero. No molecules can escape the layer in which they sublimated. Therefore, only the number of molecules can sublimate or re-sublimate that is necessary to achieve the saturation pressure that corresponds to the temperature of that layer. We added an ice-free layer of $1 \,\mathrm{cm}$ and only used one type of ice per simulation (either H$_2$O or CO$_2$). This situation corresponds to an impermeable surface, which could result from sintering or compaction. We did not run these simulations for the orbit of comet 67P, but for different constant heliocentric distances for 100 days. For comparison to the model used in Section \ref{sec:results_outgassing}, we ran it also with this orbital setting. For a heliocentric distance of 1 au, Figure \ref{fig:pressures_tpm} shows the resulting maximal pressures for the last ten days of these simulations without any activity (no scaling of outgassing rate (A) and $\Phi$ scaling (B)) and the pressures reached by the new adaptation without diffusion in a depth of $1 \,\mathrm{cm}$, $3 \,\mathrm{cm}$ and $5 \,\mathrm{cm}$  (only H$_2$O simulated and only CO$_2$ simulated). Due to the retreat of the ice front in the first case, the given pressures are not for a constant depth, but for the current location of the ice front. In that time range, the ice is no longer influenced by the diurnal variations, but the numerical layers result in jumps in the pressure when one layer is depleted and the layer below is at a lower temperature. In the non-diffusivity case, it is visible that the $5 \,\mathrm{cm}$ depth is still influenced by the diurnal variation, however obviously less than the layers above. After 100 days, a steady state is reached. Several orders of magnitude higher pressures can be reached in this case compared to our model with outgassing into space. 

The resulting pressures at a depth of $1 \,\mathrm{cm}$ for varying heliocentric distances are included in Figure \ref{fig:Maximal_Actvity_Distance_pressure}, indicated by the symbols in Figure \ref{fig:Maximal_Actvity_Distance_pressure}, and we added fit function, which can be found in Appendix \ref{App:fits}. These values lie several orders of magnitude below the maximum pressures estimated by maximum temperatures. Compared to the weight of a $1 \,\mathrm{cm}$ dust layer, smaller maximum activity distances result, which lies for water around $3.1 \,\mathrm{au}$, for CO$_2$ around $8.2 \,\mathrm{au}$ and for CO around $74 \,\mathrm{au}$.  


\section{Discussion}
\label{sec:discussion}
It should be noted that we do not take into account any of the complex shape features that comet 67P shows. The obliquity results in seasonal effects and in fall-back of material onto the less-illuminated part of the nucleus. The complex morphology and binarity of comet 67P certainly influence the thermophysical behaviour, due to shadowing and self-heating, and, thus, influence the local outgassing rate and dust activity. Here, we only addressed the seasonal effect by a simulation run with permanent illumination. The formation of pits and cracks is observed at several locations on the comet and is likely connected to the activity. Also, our scaling of the rates measured by Rosetta to one square meter does not take the actual illumination intensity or duration into account and therefore our simulations can only give a rough estimates on outgassing and dust-emission rates. As we have seen in Section \ref{sec:results_outgassing}, there are many of our scenarios not matching the observations at all, from which we might still learn how cometary activity does not work. 

Regarding the influence of the temperature dependency of the thermal conductivity, no differentiation was possible because the results were very similar, as described in Section \ref{sec:results_sub:micro}. 

It is remarkable that in all cases where CO$_2$ causes the dust activity (either by pressure build-up (cases 1b, 1c) or by the depletion criteria (cases 2b, 3b)), its outgassing rate overcomes that of H$_2$O by a factor of at least 5, but sometimes also exceeding a factor 1000. This is in contradiction to the observations at comet 67P, in which CO$_2$ never exceeded the H$_2$O-outgassing rate. Only when H$_2$O depletion causes the activity (cases 2a, 3a), the ratio between the H$_2$O- and CO$_2$-outgassing rates is comparable to the observation. For these scenarios, the H$_2$O vapour pressure is always sufficient to overcome the weight of the ejected layers. In this work, we focus on pebbles as isotropic, homogeneous agglomerates and the ejection of dust in the size range of centimetres. Other models assume that pebbles possibly possess a substructure, resulting in micrometre-sized-dust activity \citep{Fulle.2019,Fulle.2020,Fulle.2020b,Fulle.2022b}. Our approach has certainly many simplifications, however their influence is mainly at the surface. The outgassing of CO$_2$ in deeper layers in case of H$_2$O-depletion-driven dust activity is less influenced by this. Our results lead us to the conclusion that H$_2$O seems to be the main driver of activity on comet 67P, as otherwise higher CO$_2$-outgassing rates would be expected. However, our dust production rate and dust-to-H$_2$O ratio always exceeds the observations. A variation of the used parameter set, e.g., the dust-to-ice mass ratio, might result in better matches, which is foreseen in future work. Our model presented here does not produce dust chunks, which are likely caused by CO$_2$ outgassing, but only ejects dust in the few centimetres scale. This is in contrast to the findings of \citet{Gundlach.2020}. Regarding the dust-size distribution around perihelion, chunks with sizes of several decimetres dominate the dust production in mass \citep{Fulle.2016c}. Despite this dominance in mass, it is likely that the ejection of chunks is not a phenomenon homogeneously distributed over the cometary surface, but discretely located, so that our simple approach cannot reproduce it and more ambitious modelling would be necessary, which is beyond the scope of this work. 

Such a detailed study of one region can be found in \citep{Davidsson.2022}. \citet{Davidsson.2022} model a location at the Hapi region before the formation of a pit occurring in late 2014 and fitted it to MIRO data. With this adaptation of the  \texttt{NIMBUSD} model, they find a change in diffusivity, which is their free parameter (expressed by radius and length of a pore tube), in October and November 2014. With the resulting parameters, they derived CO$_2$ pressures up to $2.4 \,\mathrm{kPa}$ around pit formation time. However, they assume a high inner temperature of $120 \,\mathrm{K}$, which results in a CO$_2$ backgound pressure in the interior of $\sim 4 \,\mathrm{Pa}$. In our understanding of the cohesion and the weight of the layer above it, the material at the surface should not be able to withstand pressures up to kPa. The work of \citep{Capria.2017} showed that it is unlikely that the heat wave reached depths lower than roughly one meter in the cometary lifetime when accounting for the surface erosion. They conclude that below this active layer the comet should have remained pristine and cold. This is why we used an interior temperature of $50 \,\mathrm{K}$, in contrast to a temperature of $120 \,\mathrm{K}$ assumed by \citet{Davidsson.2022}. 

In this study we used a fixed set of parameters for the cometary-material properties, which are only educated guesses and might deviate from the real values. For example, \citet{Davidsson.2021} found lower values for the dust-to-ice mass ratio to fit their thermophysical model to the Rosetta data. However, constraining these values was not the scope of this work. Here we intended to first investigate the influence of the activity mechanism and the scaling of the outgassing rate for one case assuming temperature-dependent thermal conductivity and for another case assuming a constant value. Certainly, the dust-to-ice mass ratio and the relative abundance of CO$_2$ influence the outcome of all our scenarios, as discussed before. In case of the artificial ejection mechanisms, more/less ice will result in a slower/faster erosion of dust. Regarding our results with some agreement for several scenarios, it will be interesting for future work to investigate the influence of the chosen parameters. This is especially true for the models with reduced diffusion lengths. For pressure-driven and CO$_2$-depletion-driven activity, our simulations do not match observations, however for H$_2$O-depletion-driven activity (case 2a), the slope of the H$_2$O-outgassing rate is similar to Rosetta findings. However the absolute values are clearly too low. Here, the variation of the fixed parameters might change this. 

In three scenarios, the day-night cycle was turned off and the illumination was therefore permanent. This is an extremely unrealistic case, because permanent illumination on comet 67P was only present for some part of the orbit and for some parts of the nucleus. However, with this easy approach, we can investigate whether the obliquity of the comet can explain the  mismatch between observations and modelling. We used one case with temperature-dependent thermal conductivity and two cases with a constant thermal conductivity of $10^{-4}\,\mathrm{W/(K \,m)}$ and $10^{-1}\,\mathrm{W/(K \,m)}$, respectively. The low conductivity leads to very low outgassing rates, well below the measured vales. Only the dust-production rates fall into the observed range. For the pebble case and the high thermal conductivity and for most part of the orbit, the outgassing rates exceed the observations. However, in all cases the  observed perihelion water production was not reached. Therefore, the seasonal effects on comet 67P do not explain the deficiency of our model. In general, the slope of the outgassing rates when the comet approaches perihelion was in all model variations with high diffusivity too shallow compared to the observed slope. Other effects must lead to the observed steep increase, e.g. by the decreasing depth of the sublimation fronts and changed diffusivity, as shown in our scenario with reduced diffusion length. However, the ejection of larger chunks nearer to the Sun, remains unclear when the sublimation depth shrinks. 

In the case of reduced diffusion length, i.e. $b = 10\,\mathrm{\mu m}$, the achieved pressures were a lot higher, reaching $0.1 \,\mathrm{Pa}$ expected for the tensile strength of a pebble packing. To investigate the maximum pressures that can be reached, we used an adaptation of our model without actual outgassing of ices into space, which is equivalent to an impermeable structure. In this case, the pressure can be several orders of magnitude higher and can even reach MPa, which is mostly unreasonable for comets, because of their high outgassing rates. However, if locally the gas is trapped by an impermeable layer, which might form from sintering or compaction, locally higher pressures in the kPa range might be achievable. Besides the constraints of maximum activity distance shown in Section \ref{sec:results:pressure}, we can conclude that pressures exceeding the tensile strength estimations might be possible. However, it should be noted that the diffusivity must be small in all spatial directions, otherwise gas would escape laterally without the build-up of such pressures. 

\section{Conclusions}
\label{sec:conclusions}
In this study we investigated the influence of different model assumptions on the outgassing and dust-production rates of a comet. 

The comparison between a temperature-dependent and a constant thermal conductivity showed no significant difference. Hence, from this work we cannot constrain this aspect and can conclude that the influence of the temperature dependency is minor. 

We also investigated the influence of a scaling of the outgassing rate of H$_2$O and CO$_2$ by the local volume filling factor and compared it to the no-scaling case. Also here, we have only seen minor differences. The volume-filling-factor scaling resulted in slightly higher pressures (see Figure \ref{fig:pressures_tpm}), but in our scenarios this would only influence the results when placing the tensile strength value in the range between $10^{-2} \,\mathrm{Pa}$ and $10^{-1} \,\mathrm{Pa}$, which we did not further investigate.

Additionally, we compared different dust-activity mechanisms. Beside the before-used mechanism, where layers get ejected when a pressure threshold is reached by the vapour pressure of one of the ices, we introduced artificial activity mechanisms. Here, either a fixed number of layers gets ejected as soon as they are depleted of H$_2$O (case 2a) or CO$_2$ (case 2b) or once per cometary day all ice-free layers get ejected.  

It should be kept in mind that we do not take into account any of the complex shape features that comet 67P shows. Shadowing and self-heating can influence the local outgassing rate and dust activity. We only addressed the seasonal effect by a simulation run with permanent illumination. We scaled the rates measured by Rosetta to one square meter and did not take the actual illumination intensity or duration into account. Therefore, they can be interpreted as lower limits in comparison to our simulations of an equatorial square meter.

To constrain the maximum pressures reachable in comets, we performed simulations with an adapted model, equivalent to an gas-impermeable dust structure. With this, we cannot investigate outgassing rates, as no gas escapes into space. 

In summary, our main results are:
\begin{itemize}
    \item If CO$_2$  pressure or depletion is responsible for the activity, it dominates the outgassing rate and exceeds that of water, even though CO$_2$ ice is less abundant. Only in the artificial case, where water depletion  is responsible for the activity, the ratio of the outgassing rates falls into the observed range. 
    \item In case of CO$_2$-depletion-driven activity, it can result in the ejection of more than one pebble layer, which contains water ice. H$_2$O-depletion-driven activity emits only pebble-sized dust. 
    \item With a diffusion length of $b = 10 \,\mathrm{mm}$, expected for a pebble packing, we cannot match the increasing slope of the outgassing rate vs. heliocentric distance. Our results are in any scenario shallower than the measured curves. When decreasing the diffusion length to $b = 10 \,\mathrm{\mu m}$, the slope is similar to the observations, but the absolute H$_2$O-outgassing rates are too low. Even for permanent illumination, the observed perihelion outgassing rates cannot be reached by our model. 
    \item For the pressure-driven activity mechanism, the diffusion length only influences the maximally-reachable pressure. If the chosen threshold is overcome (in all cases by the CO$_2$ pressure), the resulting gas and dust production rates are not influenced by the diffusion length. 
    \item Pressures to overcome expected tensile strength for pebble piles or the weight of meter-sized boulders are only reached with reduced diffusivity ($b = 0 ... 10 \,\mathrm{\mu m}$) compared to a packing of pebbles without additional fluffy dust filling the voids \citep{Fulle.2017}. Additionally, a low outgassing rate is necessary to reach high enough temperatures for such pressures, because otherwise too much energy is consumed by latent heat of sublimation.
    \item In our model in which no gas is transported through the dust layers, pressures exceeding the weight of a $1 \,\mathrm{cm}$ dust layer on 67P (see Figure \ref{fig:Maximal_Actvity_Distance_pressure}) can be reached for H$_2$O inside $\sim 3.1 \,\mathrm{au}$, for CO$_2$ inside $\sim 8.2 \,\mathrm{au}$ and for CO inside $\sim 74 \,\mathrm{au}$. Those distances can be interpreted as maximum activity distances for the respective species. 
\end{itemize}

To really understand the dust-ejection mechanism in comets, advanced modelling needs to be connected to laboratory work and comet observations. These models might take into account effects of sintering, thermal cracking and other phenomena, which we here have not investigated.

\section*{Acknowledgements}

This work was funded through the DFG project BL 298/27-1. N.A.’s contributions were made in the framework of a project funded by the European Union’s Horizon 2020 research and innovation programme under grant agreement No 757390 CAstRA. We thank Thilo Glißmann for performing simulations of condensation of molecules in a pebble structure. We also thank Christopher Kreuzig for providing us with information about the outgassing behaviour of granular water ice.

\section*{Data Availability}
The data underlying this article will be shared on reasonable request to the corresponding author.




\bibliographystyle{mnras}
\bibliography{paper_references} 


\appendix

\section{Details Of The Thermophysical Model}
\label{App:TPM}
The parameters used in this study are chosen to fit observations of comet 67P/Churyumov-Gerasimenko and can be found in Table \ref{Tab:1_Parameters}. 

For the mass-density assumption, this means that the total apparent density of the nucleus is known with high accuracy \citep{Jorda.2016}. However, the density of the solid material, e.g., the dust and ice grains, is not known very well as the composition and the porosity has a large uncertainty \citep[see e.g. Section 5.1.3 and Figure 2 in][]{Blum.2022}. To fit the total nucleus density with our assumptions about the dust-to-ice mass ratio and the porosity, we assumed a corresponding dust density of $3,645 \,\mathrm{kg/m^3}$

Also, the specific heat capacity and the thermal conductivity of the grains (which are used to calculate the total thermal conductivity in the pebble case, see Equation \ref{eq:lambda_net_macro}), depend on the composition and are calculated with respect to the mass fraction of the materials. The specific heat capacity is temperature dependent. For water ice the formula 
\begin{equation}
    c_{p,\mathrm{H_2O}}(T) = 7.5 \, T + 90 \, \mathrm{[J/(kg \, K)]}
    \label{eq:heat_capacity_water}
\end{equation}
from \citet{Klinger.1981} was used. For CO$_2$ ice, a fit to the data of \citet{Giauque.1937} was used, which reads
\begin{multline}
    c_{p,\mathrm{CO_2}}(T) = -291.24 + 24.17 \, T - 0.161 \, T^2 + 4.03 \times 10^{-4} \, T^3 \\ \mathrm{[J/(kg \, K)]} .
    \label{eq:heat_capacity_carbondioxid}
\end{multline}
For the refractory part, we assumed the heat capacity measured by \citet{Opeil.2020} for carbonaceous chondrites (Cold Bokkeveld),
\begin{equation}
    c_{p,\mathrm{dust}}(T) =  G + H \, T + J \, T^2 - L \, T^3 + M \, T^4 \, \mathrm{[J/(kg \, K)]} ,
    \label{eq:heat_capacity_dust}
\end{equation}
with the fit parameters $G=2.168 \times 10^{-1}$, $H=42.581 \times 10^{-2}$, $J=4.425 \times 10^{-2}$, $L=2.06 \times 10^{-4}$ and $M=2.853 \times 10^{-7}$.

The saturation pressure $p_{\mathrm{sat},X}$ and the latent heat $\Lambda_X$ of species $X$ are temperature-dependent \citep{Huebner.2006}. They are calculated with the parameters from Table \ref{Tab:1_Parameters2} with
\begin{equation}
    \Lambda_X(T) = \left( -b_X \ln(10) + (c_X-1) T + d_X \ln(10) \, T^2 \right) \frac{R_g}{m_{\mathrm{mol},X}}   \, \mathrm{[J/kg]}, 
    \label{eq:latent_heat_T}
\end{equation}
with the gas constant $R_g$ and 
\begin{equation}
    \log_{10} p_{\mathrm{sat},X}(T) = a_X + b_X/T + c_X \log_{10} T + d_X \, T \, \mathrm{[Pa]}.
    \label{eq:pressure_sat_T}
\end{equation}

As described in Section \ref{sec:TPM:Basis}, the thermal conductivity of our pebble case consists of two parts, a network part and a radiative part (see Equation \ref{eq:pebble_conductivity}). In the following, we briefly describe the model behind it. 

The radiative thermal conductivity can be written as
\begin{equation}
    \lambda_\mathrm{rad}(T) = \frac{16}{3} \sigma \, l \, T^3 ,
\end{equation}
with the Stefan-Boltzmann constant $\sigma$ and the mean free path $l$ inside the voids. 
The latter depends on the pebble radius $R$ and on the volume filling factor of the pebble packing $\Phi_\mathrm{pack}=0.6$, 
\begin{equation}
    l = e R \frac{1-\Phi_\mathrm{pack}}{\Phi_\mathrm{pack}}.
\end{equation}
The scaling factor is $e=1.34$ and was empirically estimated by \citet[][]{Gundlach.2012}. The radiative energy transport inside the pebbles is neglected, because of the small mean free path between the grains, which are the small particles (radius $r$) of which the pebbles consists.  

For the network conductivity in a granular medium, the Hertz factor describes the reduced contact area between the grains. The network thermal conductivity reads
\begin{equation}
    \lambda_\mathrm{net, micro} = \lambda_\mathrm{par} \, H_\mathrm{micro},
    \label{eq:lambda_net_micro}
\end{equation}
with the material thermal conductivity $\lambda_\mathrm{par}$ depending on the composition (Equation \ref{eq:mass_fraction1}), and the Hertz factor $H_\mathrm{micro}$, which is given by
\begin{equation}
    H_\mathrm{micro} = \left[\frac{9 (1 - \mu_\mathrm{par}^2)}{4 E_\mathrm{par}} \pi \gamma_\mathrm{par} r^2 \right]^{1/3} \xi(\Phi_\mathrm{micro},r).
\end{equation}
Here, $\gamma_\mathrm{par}$, $E_\mathrm{par}$ and $\mu_\mathrm{par}$ are the specific surface energy, the Young's modulus and the Poisson ratio of a grain, respectively, and $\Phi_\mathrm{micro}$ is the volume-filling factor of the structure inside a pebble. The geometry of the packing of the particles is taken into account by the empirical factor $\xi(\Phi_\mathrm{micro},r)$, which depends on the volume-filling factor of the micro-porous structure and on the particle size \citep[][]{Gundlach.2012}, 
\begin{equation}
    \xi(\phi_{\rm micro},r) = \frac{f_1 \, \exp\left[f_2 \, \phi_{\mathrm{micro}}\right]}{r} \, .
\end{equation}
Here, the two parameters are $f_1 = 5.18 \cdot 10^{-2}$ and $f_2 = 5.26$.

The equations above describe the network conduction inside a pebble. For a macro-porous structure built up by pebbles, the reduced contact area between the pebbles results into a further reduced network conductivity. Therefore, a second Hertz factor $H_\mathrm{macro}$ is introduced \citep[][]{Gundlach.2012}
\begin{equation}
    H_\mathrm{macro} = \left[\frac{9 (1 - \mu_\mathrm{agg}^2)}{4 E_\mathrm{agg}} \pi \gamma_\mathrm{agg} R^2 \right]^{1/3} \xi(\Phi_\mathrm{pack},R).
\end{equation}
The calculation is based on the parameters of the pebble structure. Here, $\mu_\mathrm{agg}$, $E_\mathrm{agg}$ and $\gamma_\mathrm{agg}$ are the Poisson ratio, the Young's modulus and the specific surface energy of a pebble, respectively. The parameter $\xi(\Phi_\mathrm{pack},r)$ is defined analogously to above. The specific surface energy of a pebble, $\gamma_\mathrm{agg}$ is reduced compared to a solid grain due to its porosity. A formula for it can be found in \citet{Gundlach.2012}. The network conductivity of a macro-porous pebble structure can then be described with these factors derived above as
\begin{equation}
    \lambda_\mathrm{net, macro} = \lambda_\mathrm{par} \, H_\mathrm{micro} \, H_\mathrm{macro}.
    \label{eq:lambda_net_macro}
\end{equation}

\begin{table*}
\caption{Physical parameters used in this study. To be able to compare the simulation results with Rosetta data, properties of comet 67P were chosen.}
\begin{tabular}{lcccc}
\hline
Parameter & Symbol & Value & Unit & Reference \\ \hline
Solar constant & $I_E$ & $1367 $ & $\mathrm{W \, m^{-2}}$ & - \\
Stefan–Boltzmann constant & $\sigma$ & $ 5.67 \times 10^{-8} $&$ \mathrm{W \, m^{-2} \, K^{-4}} $ & -\\
Boltzmann constant & $k$ & $1.38 \times 10^{-23} $&$ \mathrm{J \, K^{-1}}$ & - \\
Bulk density & $\rho$ & $532 $&$ \mathrm{kg \, m^{-3}}$ & \citet{Jorda.2016} \\
H$_2$O density & $\rho $ & $934 $&$ \mathrm{kg \, m^{-3}}$ & - \\
CO$_2$ density & $\rho$ & $1600 $&$ \mathrm{kg \, m^{-3}}$ & - \\
Albedo & $A$ & $0.055$ & - & \citet{Sierks.2015} \\
Emissivity & $\epsilon$ & 1 &  - & - \\
Grain radius & $r$ & $0.1 $&$ \mathrm{\mu m}$ &    \citet{Mannel.2016,Mannel.2019}    \\
Volume filling factor of pebble packing & $\Phi_\mathrm{pack}$ & $0.55$ & - & \citet{Blum.2014,ORourke.2020}     \\
Inter-pebble volume filling factor & $\Phi_\mathrm{agg}$ & $0.4$  & - & \citet{Weidling.2009}   \\
Poisson ratio of pebble & $\mu_\mathrm{agg}$  & $0.17$ & -   & \citet{Weidling.2012}   \\
Poisson ratio of particle & $\mu_\mathrm{par}$  & $0.17$ & -  & \citet{Chan.1973}    \\
Young's modulus of pebble & $E_\mathrm{agg}$  & $8.1 \times 10^{3} $&$ \mathrm{Pa}$   & \citet{Weidling.2012}   \\
Young's modulus of particle & $E_\mathrm{par}$  & $5.5 \times 10^{10} $&$ \mathrm{Pa}$ & \citet{Chan.1973}      \\
Specific surface energy of particle & $\gamma_\mathrm{par}$  & $0.1 $&$ \mathrm{J \, m^{-2}}$ & \citet{Heim.1999}      \\
Heat conductivity of particle & $\lambda_\mathrm{par}$ & $0.5 $&$ \mathrm{W \, m^{-1} K^{-1}}$ & \citet{Blum.2017} \\
Heat conductivity of H$_2$O & $\lambda_\mathrm{par}$ & $651/T $&$ \mathrm{W \, m^{-1}}$ & \citet{Petrenko.1994} \\
Heat conductivity of CO$_2$ & $\lambda_\mathrm{par}$ & $0.02 $&$ \mathrm{W \, m^{-1} K^{-1}}$ & www.nist.gov \\
Packing structure coefficient & $f_1$ & $5.18 \times 10^2$ & -  & \citet{Gundlach.2012} \\
 & $f_2$ & $5.26$ & -  & \citet{Gundlach.2012} \\
Mean free path coefficient & $e$  & $1.34$ & -  & \citet{Gundlach.2012} \\
\hline
\end{tabular}
\label{Tab:1_Parameters}
\end{table*}

\begin{table*}
\caption{Parameters for temperature-dependent saturation pressure and latent heat \citep[][]{Huebner.2006}}
\begin{tabular}{lccccc}
\hline
Species $X$ & $a_X$ & $b_X$ & $c_X$ & $d_X$ & $m_{\mathrm{mol},X}$\\
  & [-] & [K] & [-] & [$\mathrm{K}^{-1}$]  & [kg/mol] \\
\hline
H$_2$O & $4.07023$ & $-2484.986$ & $3.56654$ & $-0.00320981$ & $0.018$ \\
CO$_2$ & $49.2101$ & $-2008.01$ & $-16.4542$ & $0.0194151$ & $0.044$\\
CO & $53.2167$ & $-795.105$ & $-22.3452$ & $0.0529476$ & $0.028$ \\

\hline
\end{tabular}
\label{Tab:1_Parameters2}
\end{table*}

\section{Rosetta Data Fit Functions}
\label{App:Rosetta_Data}
\begin{figure}
    \centering
    \includegraphics[width=\columnwidth]{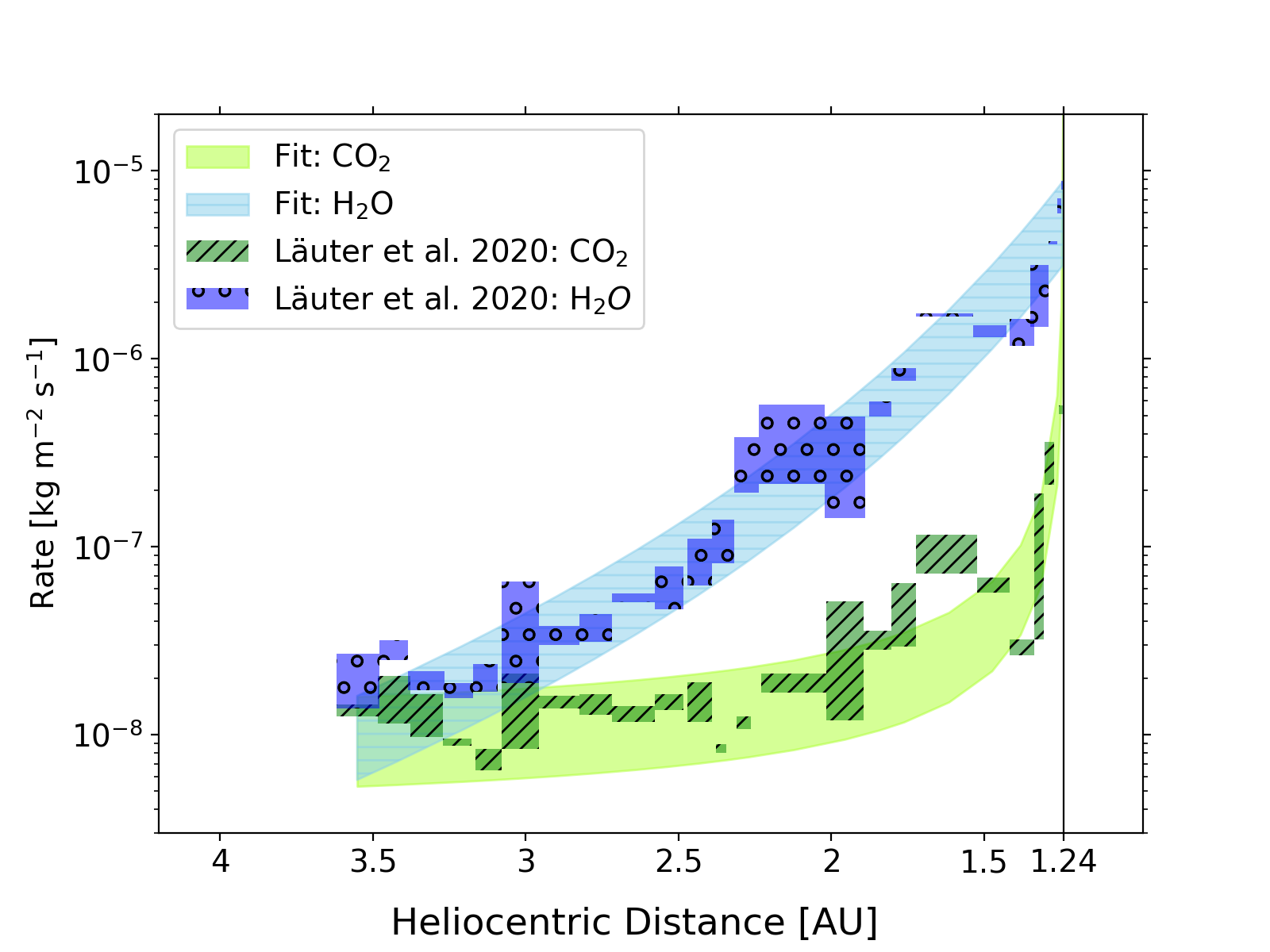}
    \caption{Comparison between outgassing rate of H$_2$O and CO$_2$ from ROSINA measurements for the inbound orbit from \citet{Lauter.2020} to our approximation functions.}
    \label{fig:Rosina_Fit}
\end{figure}

As described shortly in Section \ref{sec:results_outgassing}, we approximated the ROSINA data from \citet{Lauter.2020} with a fit function. Due to the wide and changing scattering of the data, it was not possible to achieve a single and well-converting fit. Therefore, we used parameter ranges for the functions so that much of the data is covered. Because we only 
qualitatively compare our results with this data, this approximation does not influence our findings. 

For the H$_2$O-outgassing rate, a power law reproduced the measured data best with
\begin{equation}
    q_{\mathrm{H_2O}}(r_H) = (2.2 \pm{1.0}) \times 10^{-5} \, r_H^{-6} [\mathrm{kg \, m^{-2} \, s^{-1}}].
\end{equation}
For the CO$_2$-outgassing rate, a power low did not fit the data well. Therefore, we decided to use a hyperbolic fit with
\begin{equation}
    q_{\mathrm{CO_2}}(r_H) = (6.9 \pm 3.5) \times 10^{-5} \, \frac{r_H}{r_H - 1.24 \,\mathrm{au}}[\mathrm{kg \, m^{-2} \, s^{-1}}].
\end{equation}
Figure \ref{fig:Rosina_Fit} shows the data from \citet[][]{Lauter.2020} as well as our resulting approximations.

\section{Fit functions of pressure estimation}
\label{App:fits}
In Section \ref{sec:results:pressure}, we presented an estimation for the pressures $p$ in a depth of $1 \,\mathrm{cm}$ as a function of the heliocentric distance $r$ and under the condition that no molecules can diffuse outwards. To estimate the distance at which a pressure of $10^{-3} \,\mathrm{Pa}$ can no longer be achieved by the respective ice species, we used fit functions to the data points. The resulting functions are presented below. For each species $X$, we used a function for the temperature as a function of the heliocentric distance and calculated the saturation gas pressure by
\begin{equation}
    p(r, X) = p_{\mathrm{sat}, X}(T(r)) .
\end{equation}
The resulting temperatures are for H$_2$O:
\begin{equation}
    T(r) = \left(\frac{(1-A) \, I_0}{36.55 \, \sigma \, \epsilon \, r^2} \right)^{1/3.5} ,
\end{equation}
for CO$_2$:
\begin{equation}
    T(r) = \left(\frac{(1-A) \, I_0}{0.24 \, \sigma \, \epsilon \, r^2} \right)^{1/3.3} ,
\end{equation}
and for CO:
\begin{equation}
    T(r) = \left(\frac{(1-A) \, I_0}{1.70 \, \sigma \, \epsilon \, r^2} \right)^{1/4.2} .
\end{equation}


\bsp	
\label{lastpage}
\end{document}